\documentclass[%
 reprint,
 amsmath,amssymb,
 aps,
]{revtex4-2}

\usepackage{hyperref}
\hypersetup{
    colorlinks=true,
    linkcolor=black,
    citecolor=black,
    urlcolor=black
}

\usepackage{graphicx}
\usepackage{dcolumn}
\usepackage{bm}
\usepackage{array}
\usepackage{color}

\newcommand{\Exc}{E_{\text{xc}}}
\newcommand{\Eel}{E_{\text{el}}}
\newcommand{\Eelloc}{E_{\text{el}}^{\text{loc}}}

\newcommand{\EH}{E_{\text{H}}}
\newcommand{\EZZ}{E_{\text{ZZ}}}
\newcommand{\bR}{\textbf{R}}
\newcommand{\br}{\textbf{r}}

\newcommand{\bd}{\textbf{d}}
\newcommand{\dr}{d\textbf{r}}

\newcommand{\vxc}{v_{\text{xc}}}

\newcommand{\velloc}{v_{\text{el}}^{\text{loc}}}
\newcommand{\vKS}{v_{\text{KS}}}
\newcommand{\vpsploc}{v_{\text{psp}}^{\text{loc}}}
\newcommand{\vpspnl}{v_{\text{psp}}^{\text{nl}}}
\newcommand{\vpspnlop}{\hat{v}_{\text{psp}}^{\text{nl}}}
\newcommand{\bsmear}{b_{\text{sm}}}
\newcommand{\vsmear}{v_{\text{sm}}}
\newcommand{\bu}{b_{\text{u}}}
\newcommand{\EselfZZ}{E_{\text{ZZ}}^{\text{self}}}

\newcommand{\bpsi}{\boldsymbol{\psi}}

\newcommand{\bZ}{\mathbf{Z}}
\newcommand{\bZtilde}{\mathbf{\widetilde{Z}}}
\newcommand{\bV}{\mathbf{V}}
\newcommand{\bVin}{\mathbf{V}_{\text{in}}}
\newcommand{\bVout}{\mathbf{V}_{\text{out}}}

\newcommand{\bME}{\mathbf{M}^{\text{E}}}
\newcommand{\bMEtilde}{\mathbf{\widetilde{M}}^{\text{E}}}
\newcommand{\bHE}{\mathbf{H}^{\text{E}}}
\newcommand{\bA}{\mathbf{A}}
\newcommand{\bD}{\mathbf{D}}
\newcommand{\bI}{\mathbf{I}}
\newcommand{\bG}{\mathbf{G}}
\newcommand{\bLambda}{\boldsymbol{\Lambda}}
\newcommand{\hmin}{h_{\text{min}}}
\newcommand{\hmax}{h_{\text{max}}}
\newcommand{\Ecut}{E_{\text{cut}}}

\newcommand{\supC}{^{\text{C}}}
\newcommand{\supE}{^{\text{E}}}
\newcommand{\phirem}{\phi^{\text{rem}}}

\newcolumntype{C}[1]{>{\centering\arraybackslash}p{#1}}

\begin{document}

\preprint{APS/123-QED}

\title{Large-scale pseudopotential density functional theory calculations using orthogonalized enriched finite element basis}

\author{Avirup Sircar}
\email{avirup@umich.edu}
\author{Bikash Kanungo}
\author{Sambit Das}
\author{Vikram Gavini}
\email{vikramg@umich.edu}
\affiliation{%
 Department of Mechanical Engineering, University of Michigan, Ann Arbor, MI 48109, USA
}%

\begin{abstract}
We present an efficient and scalable computational framework for pseudopotential Kohn-Sham density functional theory (KS-DFT) calculations using an enriched finite element (EFE) basis. The EFE basis is formed by augmenting the classical finite element (CFE) basis with compact atom-centered functions, which we term enrichment functions. The key idea is to combine the completeness of a finite element basis with the efficiency of an atom-centered basis. We orthogonalize the enrichment functions with respect to the underlying CFE basis to simultaneously improve the conditioning of the EFE basis and the efficiency of evaluating the inverse of the overlap matrix. To efficiently solve the Kohn-Sham eigenvalue problem, we employ a residual-based Chebyshev subspace iteration approach that is tolerant to approximations in the evaluation of the inverse of the overlap matrix. We demonstrate the accuracy of the framework as compared to the widely available DFT packages.  For benchmark non-periodic calculations, ranging up to 39,083 electrons, the EFE basis offers a  $5-7\times$ reduction in degrees of freedom over the CFE basis. As a result of this, EFE achieves a $5-9\times$ reduction in computational cost over the CFE basis. The EFE basis also provides a $4-5\times$ reduction in the required memory compared to the CFE basis, thus allowing for optimal utilization of computational resources. Finally, we demonstrate that the EFE basis affords good parallel scalability. Overall, the EFE basis offers a systematically convergent, fast, scalable, resource-efficient basis for pseudopotential DFT calculations.
\end{abstract}
\maketitle


\section{Introduction}\label{sec:intro}
Density functional theory (DFT) is the most widely used electronic structure method. It establishes all ground state properties of materials to be a functional of the ground state electron density~\cite{hohenberg1964inhomogeneous}. The practicality of the theory is achieved by the Kohn-Sham ansatz~\cite{kohn1965self}, which replaces the immensely complicated and computationally intractable interacting many-electron  Schrodinger equation with an auxiliary system of non-interacting electrons in an effective mean field. This simplification rests on the existence of a universal exchange–correlation energy ($\Exc$) which encapsulates all the quantum many-electron interactions as a functional of the electron density ($\rho(\br)$). Although the form of the exact $\Exc[\rho]$ is unknown, many useful approximations to it have been developed over the past 50 years~\cite{Cohen2012Challenges, Burke2012Perspective, Jones2015Density, Mardirossian2017Thirty, Teale2022DFT}. Beyond the XC approximation, which is unavoidable, most DFT calculations also invoke the pseudopotential (PSP) approximation~\cite{Pickett1989PSP, Chelikowsky2000PSP, schwerdtfeger2011pseudopotential, martin2020electronic}. The PSP approximation is based on two observations. First, in most cases, the core electrons are chemically inert, and only the valence electrons participate in chemical binding. Second, the core electrons screen the singular nuclear potential, due to which the valence electrons see a much smoother \emph{pseudopotential}. Given a pseudopotential, it suffices to only solve for the smooth pseudo-orbitals corresponding to the valence electrons. The PSP approximation offers two major computational advantages. First, it mitigates the need for a refined discretization otherwise needed to resolve the singular nuclear potential and the oscillatory orbitals in the core region. Second, it allows for solving for a much smaller number of electrons (valence electrons over all-electron). As a result, a vast majority of DFT codes are based on the pseudopotential approximation.

The construction of a pseudopotential is not unique. However, there are a few common considerations in almost all standard PSPs. These include matching the pseudo-orbital for the valence electron to the all-electron orbital beyond a cutoff region around the atom and matching the scattering properties of the valence electrons at atomic eigenvalues to their all-electron counterpart. Beyond these basic considerations, different PSPs include additional constraints for greater transferability. For instance, norm-conserving (NC) PSPs~\cite{hamann1979norm} also preserve the electron charge of the pseudo-orbital inside the cutoff region. However, the norm-conserving constraint often leads to harder (deeper and more oscillatory) PSPs. Ultrasoft PSPs ~\cite{vanderbilt1990soft} mitigate it by relaxing the norm-conserving criteria. Specifically, it splits the pseudo-orbital into an ultrasoft part that violates the norm-conservation and a core augmentation part to compensate for any loss of charge. The projector augmented wave (PAW) method offers a philosophically different approach (PAW)~\cite{blochl1994projector} where the all-electron orbitals within a cutoff are reconstructed through a linear transformation of the pseudo-orbitals. Both ultrasoft and PAW, on account of smoother PSPs, offer greater computational efficiency. Another widely used PSP is the Goedecker-Teter-Hutter (GTH) potential~\cite{goedecker1996separable}, which, unlike the numerical form of the NC, ultrasoft, and PAW potentials, is defined analytically in terms of Gaussian and polynomial functions. More recently, a variant of NC PSPs called the optimized norm-conserving Vanderbilt (ONCV) PSP~\cite{hamann2013optimized}, offers both computational efficiency and good transferability by combining norm-conservation with the key features of ultrasoft PSPs.

Despite decades of progress in building efficient DFT codes for PSP calculations, attaining both speed and access to large system sizes remains a steep challenge. Fast DFT calculations are a necessity for high-throughput screening of materials. At the same time, large-scale DFT calculations are essential to resolve the relevant length- and time-scale in a wide range of scientific applications. For instance, our understanding of emergent phenomena in moir\'e materials~\cite{Carr2020electronic}, the defect interactions in metals and semiconductors \cite{ismail2000ab,shin2013possible,das2017electronic,rodney2017ab,das2026}, the ion conductivity and diffusivity in solid-state electrolytes \cite{dive2018molecular}, or the complex biomolecular reactions in organisms \cite{cole2016applications} all require DFT calculations that are both accurate and computationally efficient for systems containing many thousands of atoms as well as spanning metals and insulators. However, routine DFT calculations are typically limited to systems with only a few hundred atoms. Beyond the length-scale challenge, the efficiency of state-of-the-art DFT methods for even small-to-medium scale systems leaves ample room for improvement. The computational challenge in DFT is further compounded when it comes to \textit{ab initio} molecular dynamics, as it often requires accessing long time-scales as well. This work aims to address this fundamental length-scale and efficiency challenge in DFT, within the context of PSP approximation.


At the heart of all DFT calculations lies the Kohn–Sham equation, a nonlinear eigenvalue problem whose solution provides the Kohn-Sham eigenfunctions (orbitals) and eigenvalues. Several discretization schemes have been successfully applied to solve the Kohn-Sham equation. The most popular discretization scheme in DFT with PSP calculations is the planewave basis~\cite{Payne1992}, owing to its systematic convergence. The planewave basis has been adopted in widely used DFT softwares such as \texttt{VASP}~\cite{Kresse1996efficient}, \texttt{Quantum Espresso}~\cite{Giannozzi2009quantum}, and \texttt{ABINIT}~\cite{Gonze2002first}.  However, planewaves, being periodic in nature, do not lend well to systems needing non-periodic (molecules, nanoclusters) or semi-periodic (slabs, 2D materials) boundary conditions. Further, planewaves, being extended in nature, suffer from limited parallel scalability. Another popular and highly efficient choice of discretization is the atomic orbital (AO) basis~\cite{Jensen2013atomic}. The AO basis forms the backbone of widely used DFT softwares for PSP calculations, such as \texttt{CP2K}~\cite{Kuhne2020cp2k}, \texttt{SIESTA}~\cite{Soler2002siesta}, and \texttt{CONQUEST}~\cite{Bowler2006recent}. However, the AO basis can suffer from a lack of systematic convergence, especially for metallic systems.

To address these limitations, finite-elements~\cite{White1989finite, Tsuchida1995electronic, pask2001finite, motamarri2013higher} and finite-difference~\cite{chelikowsky1994finite, Kronik2006parsec, Ghosh2017sparc} have recently emerged as two promising alternatives. Both offer a systematically convergent real-space discretization, good parallel scalability, and easily accommodate various boundary conditions to treat periodic, semi-periodic, and non-periodic systems. While finite-difference methods are typically restricted to uniform spatial resolution, the finite-element basis naturally provides adaptive spatial resolution. Recent studies \cite{das2022dft, Das2019fast} show that finite-element can outperform planewave basis for PSP calculations, especially at larger system sizes. The adaptive resolution of finite-elements also enables all-electron~\cite{motamarri2013higher,motamarri2020dft,ghosh2019all} as well as mixed all-electron and pseudopotential treatments \cite{ghosh2021}. The open-source \texttt{DFT-FE} software \cite{motamarri2020dft, das2022dft} incorporates these strengths of finite-elements and has demonstrated excellent scalability, scaling to hundreds of thousands of CPU cores and tens of thousands of GPUs, while enabling simulations of systems containing up to $O(10^5)$ electrons~\cite{das2023large}. It has been successfully applied to a wide range of large-scale DFT studies, including electronic structure calculations of large DNA molecules~\cite{zhuravel2020backbone}, quasicrystal studies~\cite{baek2025quasicrystal}, evaluation of spin Hamiltonian parameters~\cite{ghosh2019all}, hybrid DFT~\cite{subramanian2024tucker}, and inverse DFT calculations~\cite{kanungo2019exact}.

Despite these advantages, a key challenge associated with the finite-element basis is the higher number of basis functions (or degrees of freedom (DoFs)) required per atom. In particular, norm-conserving PSPs involving $d$ or $f$ electrons are often computationally demanding due to their hardness, thus requiring a larger number of basis functions. Moreover, for improved transferability, they frequently necessitate the inclusion of semicore states along with valence states, further increasing the number of valence electrons and leading to even harder pseudopotentials, especially for alkali, alkaline earth, and early transition metals. For such hard PSPs, while planewave methods can achieve chemical accuracy ($\sim10^{-4}$ Ha/atom in ground state energy) using $O(10^3)$ basis functions per atom, finite-elements require $O(10^4)$ DoFs per atom. This results in increased memory requirements for finite-element basis compared to planewaves, and also a higher computational cost in the regime of small-to-medium system sizes~\cite{motamarri2020dft}.

In this work, we alleviate the above limitations of the finite element basis by using an enriched finite element (EFE) basis. In such a basis, the standard finite element basis is augmented with compact atom-centered functions, which we term enrichment functions. For clear distinction, hereafter, we refer to the standard finite element basis as the classical finite element (CFE) basis. The key idea behind the EFE basis is to have the enrichment functions capture the sharper and rapidly oscillatory behavior of the electronic fields near the nuclei, thereby mitigating the need for a refined finite element basis. The coarser nature of the EFE basis also reduces the spectral width of the discrete Kohn-Sham Hamiltonian, and hence, accelerates any iterative eigensolver. While the efficacy of the EFE basis has been demonstrated for all-electron calculations~\cite{kanungo2017large, rufus2021fast}, its effectiveness for large-scale  PSP calculations remains unknown. This work fills this gap by developing an efficient and massively parallel framework using the EFE basis for PSP calculations. We test the capabilities of the EFE basis for different materials systems, with sizes up to 39,083 electrons, using the widely used optimized norm-conserving Vanderbilt (ONCV) PSPs. Particularly, we demonstrate a $5-9\times$ savings in computational cost and a $4-5\times$ reduction in computational memory for EFE over CFE.

The rest of the paper is organized as follows. Sec. II describes the formulation of the EFE basis. Sec. III describes the numerical aspects of solving the Kohn-Sham eigenvalue problem in the EFE basis, the most computationally expensive step in a DFT calculation. Sec. IV provides numerical experiments showcasing the accuracy, efficiency, and scalability of the EFE basis. We summarize our findings and provide our future outlook for the EFE basis in Sec. V.



\section{Background} \label{sec:back}
\subsection{Kohn Sham DFT} \label{sec:KSDFT}
We are interested in the ground-state properties of a materials system composed of $N_a$ atoms and $N_e$ electrons, where the atoms are positioned at $ \mathbf{R} = \{\mathbf{R}_1, \mathbf{R}_2, \dots, \mathbf{R}_{N_a} \} $ and have atomic numbers $Z=\{Z_1, Z_2, \ldots, Z_{N_a}\}$, respectively. In the Kohn-Sham DFT framework and within a PSP approximation, these properties are computed by solving the non-linear Kohn-Sham eigenvalue problem given by  
\begin{equation}
    \left(-\frac{1}{2}\nabla^2 + \vKS[\rho; \mathbf{R}, Z](\br) + \vpspnlop[\mathbf{R}, Z]\right)\psi_i(\br) = \epsilon_i \psi_i(\br)\,,
    \label{eq:KS}
\end{equation}
where $\epsilon_i$ and $\psi_i$ are the eigenvalues and their corresponding eigenfunctions of the Kohn-Sham Hamiltonian; $\rho$ is the electron density; $\vKS$ denotes the local multiplicative Kohn-Sham potential; and $\vpspnlop$ is the non-local PSP operator whose action on a function (say $f(\br)$) is given as $\vpspnlop f(\br) = \int \vpspnl(\br,\br')f(\br')\dr'$. Here $\vpspnl(\br,\br')$ is given as superposition of atomic non-local PSP: $ \vpspnl(\br,\br')= \sum_{I}v_{\text{psp}}^{\text{nl},I}(\br-\bR_I, \br'-\bR_I)$. 
In this work, we restrict our study to non-periodic and spin-unpolarized systems; however, extensions to periodic, semi-periodic, and spin-polarized systems can be achieved following~\cite{das2022dft}.

The electron charge density, the main quantity of interest in DFT, is given in terms of the eigenstates of Eq.~\ref{eq:KS} as:
\begin{equation}
\rho(\br)=2 \sum_{i} f\left(\epsilon_{i}, \mu\right)\left|\psi_{i}(\br)\right|^{2} \,,
\label{rhocompute}
\end{equation}
where $f(\epsilon, \mu)\in[0,1]$ is the orbital occupancy given by the Fermi-Dirac distribution, and $\mu$ is the Fermi-energy, 
determined by the conservation of the number of electrons, given by 
\begin{equation}
\int \rho(\br) \dr=2 \sum_{i} f\left(\epsilon_{i}, \mu\right)=N_{e} \,.
\label{electronconstraint}
\end{equation}
The Kohn-Sham potential, $\vKS$, in Eq.~\ref{eq:KS} is given as
\begin{equation}
\begin{split}
    \vKS[\rho; \mathbf{R}, Z](\br) &= \vxc[\rho](\br) + \velloc[\rho; \bR, Z](\br) \\
    &= \frac{\delta \Exc[\rho]}{\delta\rho}(\br) + \frac{\delta \Eelloc[\rho;\bR,Z]}{\delta\rho}(\br)\,.
\end{split}
    \label{eq:vKS}
\end{equation}
In the above, $\Exc$ is the exchange-correlation (XC) energy, with $\vxc$ denoting its functional derivative, the XC potential. The XC term accounts for the quantum many-body interactions between electrons. In this work, we have used the local density approximation (LDA) for the XC approximation, specifically, the Perdew–Wang (PW92)~\cite{perdew1992accurate} form. We note that this is only a choice and does not affect the main conclusions presented in the work. 
$\velloc$ denotes the electrostatic potential obtained as the functional derivative of the electrostatic energy $\Eelloc$, given by
\begin{equation}
    \Eelloc[\rho;\bR,Z] = \int \rho(\br) \vpsploc[\bR,Z](\br)\dr +\EH[\rho] +  \EZZ[\bR,Z]\,,
\label{eq:Eel}
\end{equation}
where $\vpsploc(\br)=\sum_{I=1}^{N_a}v_{\text{psp}}^{\text{loc}, I}(|\br-\bR_I|)$ is the superposition of the local part of PSPs from each atom; $\EH=\frac{1}{2}\int\int \frac{\rho(\br)\rho(\br')}{|\br-\br'|}\dr'\dr$ is the classical electrostatic interaction between the electrons; and $\EZZ=\sum_{I=1}^{N_a}\sum_{J>I}^{N_a}\frac{Z_IZ_J}{|\bR_i-\bR_J|}$ denotes the nuclear repulsive energy. 

The extended interactions in $\Eelloc$ can be reformulated in terms of local integrals~\cite{pask2001finite, motamarri2013higher}. To begin with, we consider a compact spherically symmetric smeared nuclear charge around the $I$-th atom, given by $\bsmear^{(I)}(\br) = -Z_I\bu(|\br-\bR_I|; r_c)$. Here $\bu(r,r_c)$ denotes a compact unit charge satisfying the properties: (i) $\bu(r > r_c) = 0$ and $4\pi\int \bu(r) r^2 \dr = 1$. In this work, we use the form of $\bu$ proposed in Ref.~\cite{pask2012linear}. If the $\bsmear^{(I)}(\br)$ from different atoms are non-overlapping, using $\bsmear(\br)=\sum_{I=1}^{N_a}\bsmear^{(I)}(\br)$, it can be shown that
\begin{equation} \label{eq:EZZbmsear}
\begin{split}
    \EZZ &= \sum_{I=1}^{N_a}\sum_{J>I}^{N_a}\int \int \frac{\bsmear^{(I)}(\br)\bsmear^{(J)}(\br')}{|\br-\br'|}\dr\dr'\\ 
    &= \frac{1}{2} \int \int \frac{\bsmear(\br)\bsmear(\br')}{|\br-\br'|}\dr\dr' - \frac{1}{2}\sum_{I=1}^{N_a}\int \int \frac{\bsmear^{(I)}(\br)\bsmear^{(I)}(\br')}{|\br-\br'|}\dr\dr' \\
    &= \frac{1}{2} \int \int \frac{\bsmear(\br)\bsmear(\br')}{|\br-\br'|}\dr\dr' - \EselfZZ \,.
\end{split}
\end{equation}
Using $\vsmear(\br)=\int\frac{\bsmear(\br')}{|\br-\br'|}\dr'$ as the electrostatic potential of $\bsmear$, we can rewrite $\Eelloc$ as 
\begin{equation} \label{eq:EellocReform}
\begin{split}
\Eelloc[\rho;bR,Z] &= \frac{1}{2}\int\int\frac{(\rho(\br) + \bsmear(\br))(\rho(\br')+ \bsmear(\br'))}{|\br-\br'|} \dr\dr' \\
&\,\,\,\,\,\,+\int \rho(\br) \left(\vpsploc(\br)-\vsmear(\br)\right) \dr - \EselfZZ\\
&= \frac{1}{2}\int (\rho(\br) + \bsmear(\br))\phi(\br)\dr  \\
&\,\,\,\,\,\,+\int \rho(\br) \left(\vpsploc(\br)-\vsmear(\br)\right) - \EselfZZ\,,
\end{split}
\end{equation}
where $\phi(\br)=\int\frac{\rho(\br)+\bsmear(\br)}{|\br-\br'|}\dr'$ is the electrostatic potential corresponding to $\rho + \bsmear$ and can be obtained from the solution of the following Poisson problem
\begin{equation} \label{eq:poisson}
-\frac{1}{4\pi}\nabla^2 \phi(\br) = \rho(\br) + \bsmear(\br)\,.
\end{equation}
The above reformulation of $\Eelloc$ is often referred to as the local reformulation and has been discussed in greater detail in~\cite{Pask1999, Pask2005, Suryanarayana2010, motamarri2012, motamarri2013higher, das2022dft}. Subsequently, the total electrostatic energy is given as $\Eel = \Eelloc + E_{\text{el}}^{\text{nl}}$, where $E_{\text{el}}^{\text{nl}}=2\sum_i f(\epsilon_i, \mu)\int \int \psi_i^{\dagger}(\br')\vpspnl(\br,\br')\psi_i(\br)\dr\dr'$.
The Kohn-Sham equations have to be solved self-consistently, upon which the ground-state energy can be computed as 
\begin{equation}
E =T_{\text{s}}+ \Eel + \Exc\,,
\end{equation}
where $T_{\text{s}}$ is the kinetic energy of non-interacting electrons given by 
\begin{equation}
\begin{aligned}
T_s=2\sum_{i=1}^{N_e} \int f(\epsilon_i,\mu)\psi_i^*(\mathbf{x})\left(-\frac{1}{2}\nabla^2\right) \psi_i(\mathbf{x})d \mathbf{x} \,.
\end{aligned}
\end{equation}
\subsection{Enriched Finite Element (EFE) Basis} \label{sec:EFEBasis}
The EFE basis, as noted earlier, is constructed by augmenting the classical finite element (CFE) basis with compact atom-centered functions, which we term enrichment functions. The computational efficiency afforded by the EFE basis results from suitably chosen enrichment functions, such as solutions to the Kohn-Sham problem for single atoms. To ensure strict locality of the EFE basis, we truncate the enrichment functions with a smooth cutoff function. This serves two purposes. First, compact enrichment functions reduce the ill-conditioning of the EFE basis, especially for large systems, as compared to more diffuse ones. Second, compact enrichment functions naturally lend to better parallel scalability. We remark that truncating the enrichment functions does not negate the benefits of enrichment functions, as they still capture the nature of electronic fields close to the nuclei, where sharp oscillations in the electronic structure can occur. Further, the EFE basis inherits the completeness of the CFE basis and hence, offers systematic convergence. Despite these advantages of the EFE basis, it can still be ill-conditioned, especially while using a refined CFE basis, where the
enrichment functions become close to being linearly dependent
on the CFE basis. This is alleviated by orthogonalizing the enrichment functions with respect to the underlying CFE basis, following the procedure introduced in Ref.~\cite{rufus2021fast}. While in the past we have referred to the resulting basis as orthogonalized-EFE (OEFE) basis, here we simply refer to it as the EFE basis. This orthogonalization procedure also
provides another advantage of rendering the overlap matrix block diagonal, which substantially aids in the evaluation of its inverse, as discussed later.

In the past~\cite{kanungo2017large, rufus2021fast, kanungo2023TDDFT}, we demonstrated the efficiency of the EFE basis for both all-electron DFT and time-dependent density functional theory (TDDFT) calculations. In particular, we showed the EFE basis to offer $100\times$ or higher speedup over the CFE basis. However, the efficacy of the EFE basis for PSP DFT calculations remains unclear, which is the focus of this study. We note that a related concept of partition-of-unity finite-element method (PUFEM)~\cite{Melenk1996,Babuska1997} has, previously, been employed for PSP DFT calculations and have been shown to be beneficial over the CFE basis~\cite{pask2017partition}. However, the work is limited to only small systems, and hence, does not probe the efficacy of any enrichment scheme for large systems. 


The EFE basis used in this study is constructed as follows. First, we obtain the pristine enrichment functions given as the solution to the Kohn-Sham problem for isolated atoms. We note that, given the spherical symmetry of isolated atoms, the Kohn-Sham eigenfunctions can be written as a radial function times a spherical harmonic. The radial part is obtained inexpensively from a 1D radial problem. These pristine enrichment functions can be precomputed and stored for each choice of pseudopotential across the periodic table. To ensure locality of the enrichment functions, we multiply the radial part by a smooth cutoff function that has the following properties
\begin{align} \label{eqnCutoffProperties}
h(r,r_0,t) = 
\begin{cases} 
1 & 0 \leq r < r_0, \\
    0 \leq  h < 1  & r_0 < r \leq r_0 + \frac{r_0}{t}\,,\\
    0 &   r > r_0 + \frac{r_0}{t}\,.
\end{cases}
\end{align}  
We use the form of $h(r, r_0, t)$ used in Ref.~\cite{kanungo2017large}. In the above, $t$ controls the smoothness of the cutoff functions, which we choose to be a fixed value of $0.33$ in this work, determined from numerical experiments on small systems. For a fixed $t$, the value of $r_0$ controls the compact support of the enrichment function. The value of $r_0$ for each enrichment function is chosen such that: (i) it captures the last turning point in the radial solution of the eigenstate; (ii) it satisfies the condition that the enrichment function integrates to unity within a tolerance of $5\times10^{-3}$; and (iii) it is located in the tail region of the single-atom Kohn-Sham orbital (i.e., the absolute value of the radial derivative of the enrichment function at $r_0$ is less than $5\times10^{-3}$). We limit the maximum value of $r_0$ to 10 a.u. to bound the overall extent of the basis. We note that the form of enrichment functions adopted here (i.e., single-atom Kohn-Sham eigenfunctions with a smooth cutoff function) is a convenient choice, and we do not claim optimality. Further, we orthogonalize the enrichment functions with respect to the underlying CFE basis. Putting all the above together, an enrichment function belonging to atom $I$ takes the following form
\begin{equation}
    N_{k,I}\supE(\br) =
    \underbrace{
    g_{nl}(r)h(r,r_0,t)\mathcal{Y}_{lm}(\vartheta,\varphi)
    \vphantom{\sum_{j\in S_{k,I}} c_j N_j^C(\mathbf r)}
    }_{\text{atomic-part}}
    -
    \underbrace{
    \sum_{j \in S_{k,I}} c_j^{k,I} N_j\supC(\br)
    }_{\text{orthogonalizing-part}}\,,
\end{equation}
where $\mathcal{Y}_{lm}$ is the real form of spherical harmonics;  $k\equiv(n,l,m)$ is composite index of the principal ($n$), angular ($l$), and magnetic ($m$) quantum numbers; and $(r,\vartheta, \varphi$) are the spherical coordinates of $\br$ with $\bR_I$ as the origin. In the above equation, the atomic part is simply the product of a single atom Kohn-Sham eigenfunction with the smooth cutoff function. The orthogonalizing part represents the component of the atomic part which lies along the CFE basis and has to be subtracted to ensure that $N_{k,I}\supE$ is orthogonal to the CFE basis.  We note that the index $j$ runs over only a small subset of the CFE basis functions, denoted by $S_{k,I}$, which typically includes the CFE basis functions that overlap with the atomic-part. We refer to~\cite{rufus2021fast} for the details on obtaining the $c_j^{k,I}$ coefficients above.

Having constructed the enrichment functions, we can represent the discrete Kohn-Sham eigenfunctions ($\psi_i^h$) in the EFE basis as
\begin{equation}
    \psi_i^h(\br) = \sum_{j=1}^{n_{h}} N_j\supC(\br)\psi_{i,j}\supC+\sum_{I=1}^{N_a}\sum_{k=1}^{n_I}N_{k,I}\supE(\br)\psi_{i,k,I}\supE,
\end{equation}
where the superscript $h$ denotes discretized fields and superscripts $\text{C}$ and $\text{E}$ distinguishes the classical finite-element basis from the enrichment functions.  Here, $N_j\supC$ is the $j$-th CFE basis function with its coefficient $\psi_{i,j}\supC$ and $N_{k,I}\supE$ is the $k$-th enrichment function associated with atom $I$ with its coefficient $\psi_{i,k,I}\supE$. Each atom contributes $n_I$ enrichment functions. We denote the total number of degrees of freedom as $M=n_h + \sum_{I=1}^{N_a} n_I$.

Using the above EFE representation of $\psi_i$ in the Kohn-Sham equation (Eq.~\ref{eq:KS}) results in the following discrete eigenvalue problem
\begin{equation} \label{eq:lineigensolveghep}
 \mathbf{H}\supE \bpsi_{i}=\epsilon_{i}\mathbf{M}\supE \bpsi_{i}\,,
\end{equation}
where $\mathbf{H}\supE$ and $\mathbf{M}\supE$ are the discrete Kohn-Sham Hamiltonian matrix and overlap matrix in the EFE basis; $\epsilon_{i}$ denotes the $i$-th discrete Kohn-Sham eigenvalue; and $\bpsi_{i}=\big\{\{\psi_{i,j}\supC\}_{j=1,\ldots,n_{h}};\{\psi_{i,k,I}\supE\}_{k=1,\ldots,n_I;I=1,\ldots,N_a}\big\}$ denotes the corresponding eigenvector containing the expansion coefficients $\psi_{i, j}\supC$ and $\psi_{i,k,I}\supE$. We note that both $\mathbf{H}\supE$ and $\mathbf{M}\supE$ matrices have a $2 \times 2$ block structure, given by
\begin{eqnarray}
& \mathbf{H}\supE=\left[\begin{array}{c|c}
\mathbf{H}^{\text{cc}} & \left(\mathbf{H}^{\text{ec}}\right)^{T} \\
\hline \mathbf{H}^{\text{ec}} & \mathbf{H}^{\text{ee}}
\end{array}\right] \,, \label{eq:Discrete_Eig_Gen_H} \\
& \mathbf{M}\supE=\left[\begin{array}{c|c}
\mathbf{M}^{\text{cc}} & \mathbf{0} \\
\hline \mathbf{0} & \mathbf{M}^{\text{ec}}
\end{array}\right] \,, \label{eq:Discrete_Eig_Gen_M}
\end{eqnarray}
where $\mathbf{H}^{\text{cc}}$ and $\mathbf{M}^{\text{cc}}$ are the classical-classical blocks which comprise matrix elements involving only the CFE basis functions. $\mathbf{H}^{\text{ec}}$  is the enriched-classical block containing the cross-term matrix elements involving both CFE basis functions and enrichment functions. We note that the $\mathbf{M}^{\text{ec}}$ block in the overlap matrix is zero as the enrichment functions are explicitly orthogonalized with respect to the CFE basis functions. Thus, $\mathbf{M}\supE$ has a block diagonal structure. $\mathbf{H}^{\text{ee}}$ and $\mathbf{M}^{\text{ee}}$ are the enriched-enriched blocks comprising matrix elements involving only the enrichment functions. 
The generalized eigenvalue problem in Eq.~\ref{eq:lineigensolveghep} can be transformed to a standard eigenvalue problem as
\begin{equation}\label{eq:lineigensolvesep}
 (\mathbf{M}\supE)^{-1}\mathbf{H}\supE \bpsi_{i}=\epsilon_{i}\bpsi_{i}\,,
\end{equation}
which we adopt in the numerical solution of the Kohn-Sham eigenvalue problem in the EFE basis.

\section{Numerical Aspects}\label{sec:numerics}
We now present the numerical aspects that crucially influence the accuracy, robustness, and computational efficiency of the EFE basis. Particularly, we will discuss: (i) an efficient approach to evaluate the inverse of the overlap (i.e., $(\mathbf{M}\supE)^{-1}$) that allows the use of the standard form of the eigenvalue problem; (ii) a subspace iteration approach to solve the Kohn-Sham eigenvalue problem based on a \emph{residual} formulation of Chebyshev filtering, which is more robust to approximations made in $(\mathbf{M}\supE)^{-1}$; and (iii) an efficient reformulation of electrostatic interactions that provide the desired accuracy using only a coarse CFE basis (i.e., without any enrichment). We discuss each aspect below.     
%
\subsection{Evaluation of Overlap Matrix Inverse}\label{sec:Overlap}
The standard form of the eigenvalue problem in Eq.~\ref{eq:lineigensolvesep} is desirable as it offers more efficient and robust eigensolvers in comparison to the generalized form in Eq.~\ref{eq:lineigensolveghep}. Importantly, the Chebyshev filtered subspace iteration approach~\cite{zhou2006self}, which has been widely adopted in many real-space DFT codes~\cite{zhou2014chebyshev,Ghosh2017sparc,motamarri2020dft} is applicable only to the standard form. In this regard, an efficient evaluation of $(\mathbf{M}\supE)^{-1}$ is crucial. As discussed earlier, $\mathbf{M}\supE$ has a block diagonal $2\times2$ block structure (cf.~Eq.~\ref{eq:Discrete_Eig_Gen_M}), allowing us to write its inverse as
%
\begin{eqnarray}
(\mathbf{M}\supE)^{-1}=\left[\begin{array}{c|c}
(\mathbf{M}^{\text{cc}})^{-1} & \mathrm{0} \\
\hline \mathrm{0} & (\mathbf{M}^{\text{ee}})^{-1}
\end{array}\right] \,.
\end{eqnarray}
An efficient evaluation of $(\mathbf{M}\supE)^{-1}$ entails an efficient evaluation of both $(\mathbf{M}^{\text{cc}})^{-1}$ and $(\mathbf{M}^{\text{ee}})^{-1}$. To simplify the evaluation of $(\mathbf{M}^{\text{cc}})^{-1}$, we use spectral finite-elements along with reduced-order Gauss-Lobatto-Legendre (GLL) quadrature rule, the combination of which results in $\mathbf{M}^{\text{cc}}$ being diagonal, thereby vastly simplifying the evaluation of $(\mathbf{M}^{\text{cc}})^{-1}$. The sufficiency of the GLL quadrature rule for evaluating the overlap matrix corresponding to spectral finite-elements has been established in Ref.~\cite{motamarri2013higher}.  

We now discuss the evaluation of $(\mathbf{M}^{\text{ee}})^{-1}$. A direct evaluation of $(\mathbf{M}^{\text{ee}})^{-1}$, while possible for small systems, can become prohibitively expensive for larger systems due to cubic-scaling complexity with the number of enrichment functions and can become the rate-limiting step in the solution of the Kohn-Sham problem. Further, the memory requirement scales quadratically with the number of enrichment functions, which necessitates parallelizing this matrix for larger systems that, in turn, can have adverse consequences for overall parallel scalability. Thus, we adopt an approach that preserves the computational efficiency and scalability without affecting the accuracy of the solution of the Kohn-Sham eigenvalue problem. If the enrichment functions belonging to each atom are arranged contiguously, $\mathbf{M}^{\text{ee}}$ will have a diagonally dominant block-structure, as enrichment functions from the same atom will have a greater spatial overlap than enrichment functions from two different atoms. Using this insight, we define an atom-block-diagonal approximation to $(\mathbf{M}^{\text{ee}})$, denoted by $(\mathbf{\widetilde{M}}^{\text{ee}})$ and given as
\begin{equation}
\mathbf{\widetilde{M}}^{\text{ee}}=
\begin{bmatrix}
    \mathbf{M}^{\text{ee}}_1  &  & &  & &\\
     & \mathbf{M}^{\text{ee}}_2 & &  & &\\
   &   &  &  & \ddots   \\
   &  & &  &   & & \mathbf{M}^{\text{ee}}_{N_a}\\
\end{bmatrix}
\end{equation}
where $(\mathbf{M}^{\text{ee}}_I)$ is overlap matrix of the enrichment functions belonging to the $I$-th atom. Since $\mathbf{\widetilde{M}}^{\text{ee}}$ has a block-diagonal form, so does its inverse, that is
\begin{equation}
(\mathbf{\widetilde{M}}^{\text{ee}})^{-1}=
\begin{bmatrix}
    (\mathbf{M}^{\text{ee}}_1)^{-1}  &  & &  & &\\
     & (\mathbf{M}^{\text{ee}}_2)^{-1} & &  & &\\
   &   &  &  & \ddots   \\
   &  & &  &   & & (\mathbf{M}^{\text{ee}}_{N_a})^{-1}\\
\end{bmatrix}
\end{equation}
$(\mathbf{\widetilde{M}}^{\text{ee}})^{-1}$ can be computed inexpensively, as it involves the evaluation of the inverse of $N_a$ small and independent matrices. Using $(\mathbf{\widetilde{M}}^{\text{ee}})^{-1}$ as an approximation to $(\mathbf{M}^{\text{ee}})^{-1}$, we define the approximate inverse of the overlap matrix as 
\begin{eqnarray} \label{eq:MEInvApprox}
(\mathbf{M}\supE)^{-1}\approx (\mathbf{\widetilde{M}}\supE)^{-1}=\left[\begin{array}{c|c}
(\mathbf{M}^{\text{cc}})^{-1} & \mathrm{0} \\
\hline \\[-2ex]
\mathrm{0} & (\mathbf{\widetilde{M}}^\text{ee})^{-1}
\end{array}\right] 
\end{eqnarray}
While $(\mathbf{\widetilde{M}}\supE)^{-1}$ offers a computationally cheap approximation to $(\mathbf{M}\supE)^{-1}$, its use can affect the accuracy of the solution to the Kohn-Sham eigenproblem. To that end, we show in the next section that one can mitigate the effect of an inexact matrix (i.e., $(\mathbf{\widetilde{M}}\supE)^{-1}$) by using the residual version of Chebyshev filtering as an iterative eigensolver.

\subsection{Residual Chebyshev Filtering}
The Chebyshev filtering-based subspace iteration approach~\cite{zhou2006self} has been effectively used in real-space DFT codes~\cite{zhou2006parallel, motamarri2013higher, motamarri2020dft, das2022dft, Ghosh2017sparc} and has been instrumental in attaining both efficiency and scalability. The central idea in Chebyshev filtering is two fold: (i) construct a good approximation to the subspace spanned by the wanted (low-lying) eigenstates of the discrete Kohn-Sham Hamiltonian, $\bA= (\bME)^{-1}\bHE$ (see Eq.~\ref{eq:lineigensolvesep}); (ii) project the Kohn-Sham eigenvalue problem onto this subspace and solve a smaller dimensional problem. 
It relies on two key properties of the Chebyshev polynomial $T_m(x)$ of degree $m$: (i) $T_m(x)$ grows rapidly for $|x|>1$, and (ii) $|T_m(x)| \le 1$ for $x \in [-1,1]$. Let the unwanted spectrum of $\bA$ lie in $(a,b)$ and let the lowest eigenvalue of $\bA$ be $\epsilon_{\text{min}}$. We define $\mathcal{L}(x)=\frac{x-c}{e}$, where $c=(a+b)/2$ and $e=(b-a)/2$ as the linear transform that maps the unwanted eigenspectrum of $\bA$ to $[-1,1]$. We incorporate this linear transformation by defining $C_m(x) = T_m(\mathcal{L}(x))/T_{m}(\mathcal{L}(a_{\text{min}}))$, where $a_{\text{min}} \leq \epsilon_{\text{min}}$ is an approximation to the lowest eigenvalue of $\bA$. In $C_m(x)$ , the factor $T_{m}(\mathcal{L}(a_{\text{min}}))$ is purely a convenient scaling choice to ensure $C_m(x) \leq 1$ when applying it to the spectrum of $\bA$ to avoid numerical overflow issues. Now, consider a trial subspace (set of vectors) $\bVin \in \mathrm{R}^{M\times N}$, where $N > N_e/2$ (typically $N \sim 1.2 \times N_e/2$). The output set of vectors $\bVout = C_m(\bA)\bVin$ represents a subspace rich in the wanted eigenspace of $\bA$. This is owing to the growth property of $C_m(x)$ (inherited from $T_m(x)$) which amplifies the components of $\bVin$ along the wanted eigenspectrum of $\mathbf{A}$. We term the construction of $\bVout$ as the Chebyshev filtering. Having obtained $\bVout$, we can approximate $\bpsi_i$ (see Eq.~\ref{eq:lineigensolvesep}) to lie in the space spanned by $\bVout$, that is,  $\bpsi_i \approx \bVout \bd_i$, where $\bd_i$ is given by the solution to the following projected eigenvalue problem
\begin{equation} \label{eq:projectedKSEig}
    \bVout^{\dagger} \bHE \bVout \bd_i = \epsilon_i \bVout^{\dagger} \bME \bVout \bd_i\,.
\end{equation}
The above approximation of $\bpsi_i$ in terms of $\bVout$, including the projected eigenvalue problem, is often referred to as the Rayleigh-Ritz procedure. Typically, for a fixed $\bA$, we perform a multi-pass Chebyshev filtering. That is, we repeat the evaluation of the Chebyshev subspace and Rayleigh-Ritz procedure with the $\bVin$ for the next iteration set to $\bVout\bD$ ($\bD$ being the collection of $\bd_i$ vectors) of the current iteration. This is continued until the eigen-residual: $||\bHE\bpsi_i - \epsilon_i\bME\bpsi_i|| < \tau, \forall i=1,2,..$, where  $\tau$ is a user-defined tolerance.

\begin{figure}[htbp]
\includegraphics[width=8.5cm]{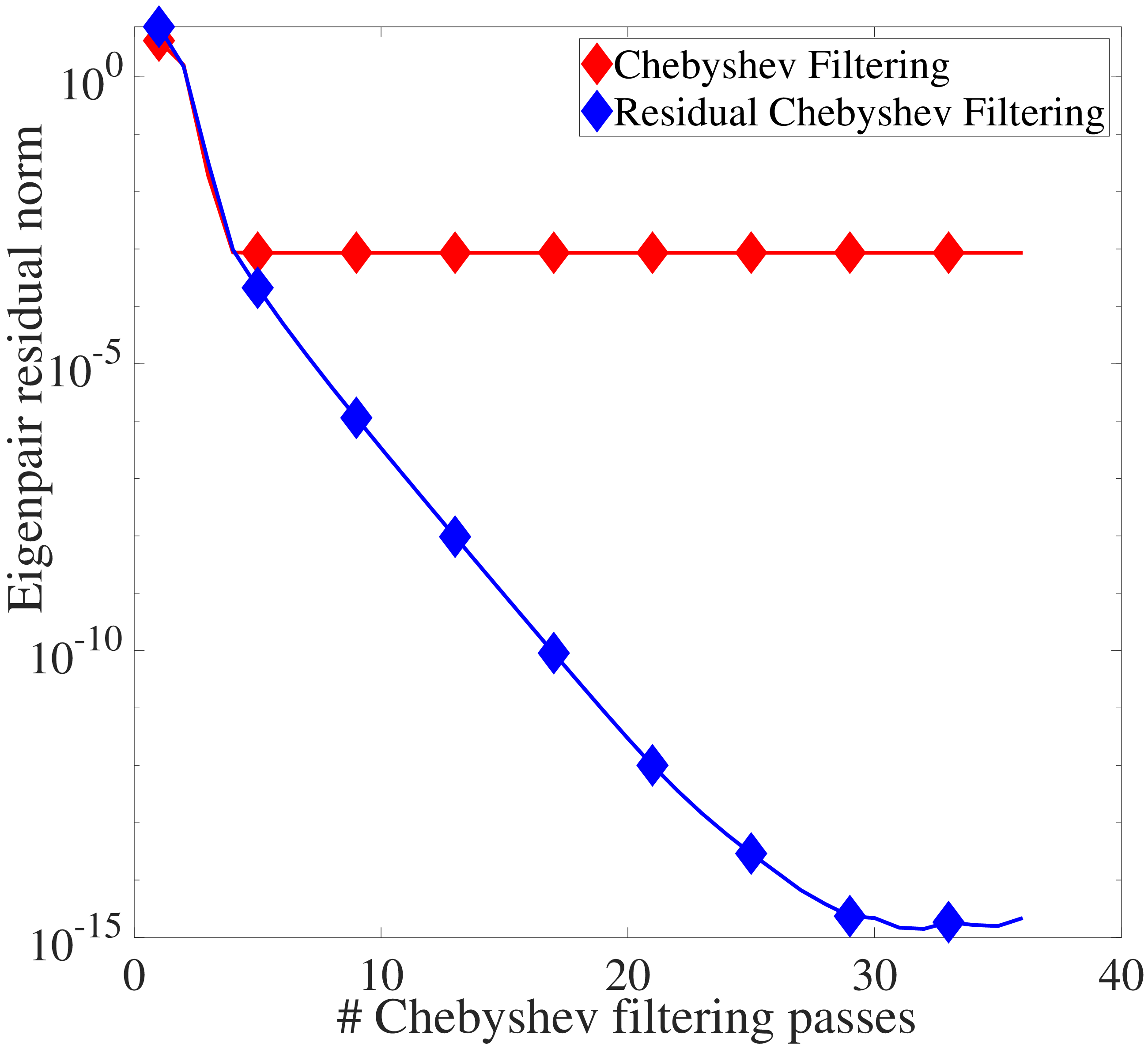}
\caption{\label{fig:epsart} Comparison of standard and residual Chebyshev filtering methods with approximate overlap matrix inverse.}
\label{fig:chebycomparison}
\end{figure}

We note that $C_m(x)$ satisfies the recurrence relation 
\begin{equation} \label{eq:ChebyRecur}
C_m(x)=\frac{2\sigma_m}{e}xC_{m-1}(x) - \frac{2c\sigma_m}{e} C_{m-1}(x)-\sigma_{m-1}\sigma_m C_{m-2}(x)\,,
\end{equation}
where $\sigma_m=1/\left(\frac{2}{\sigma_1}-\sigma_{m-1}\right)$ and $\sigma_1=e/\left(a_{\text{min}}-c\right)$. Thus, the Chebyshev filtering only requires matrix-vector products of the form $\bA\bV = (\bME)^{-1}\bHE\bV$. However, the evaluation of $(\bME)^{-1}$ or the action of $(\bME)^{-1}$ can be computationally prohibitive. To that end, we use  $(\bMEtilde)^{-1}$ (see Eq.~\ref{eq:MEInvApprox}) as an inexpensive approximation to $(\bME)^{-1}$ only for the purpose of Chebyshev filtering. We still use $\bME$ in the Rayleigh-Ritz procedure in Eq.~\ref{eq:projectedKSEig}. While this offers efficiency, the use of $(\bMEtilde)^{-1}$ can affect the robustness and accuracy of the Kohn-Sham eigensolve. To illustrate this, we consider the multi-pass Chebyshev filtering for a copper (Cu) atom. As evident from Figure~\ref{fig:chebycomparison}, the eigen-residual for the deepest Kohn-Sham eigenfunction stagnates at $8\times10^{-4}$. This stagnation is a result of approximating $(\bME)^{-1}$ with $(\bMEtilde)^{-1}$. We emphasize that the eigen-residual stagnates at a value that can affect the solution of the Kohn-Sham SCF.

In order to address the above issue, we employ a residual formulation of the Chebyshev filtering method proposed in~\cite{kodali2025residual}, which is robust to inexact matrix-vector products, which in the present context represents the inexact evaluation of $(\bME)^{-1}\bHE\bV$ in Chebyshev filtering. The main idea is to reformulate the Chebyshev filtering in terms of the eigenpair residuals. As residuals are smaller compared to the eigenvectors, the subspace iteration is more tolerant to inexact matrix-vector products. 
Following ~\cite{kodali2025residual}, we define the residual, denoted by $\mathbf{Z}$,  at the $k^{th}$ Chebyshev filter recursion step as 
\begin{equation} \label{eq:residual}
\bZ_k 
= \bMEtilde\left(
C_k\left((\bME)^{-1}\bHE\right)\bVin  
- \bVin\,C_k\left(\bLambda\right)\right)
\end{equation}
where  $\bVin$ and $\bLambda$ are the approximate eigenvectors and diagonal eigenvalue matrix obtained from the previous pass of Chebyshev filtering. Note that as $\bVin$ and $\bLambda$ approach the eigenvectors and eigenvalues of $(\bME)^{-1}\bHE$, $\bZ_k$ tends to zero. Using the recurrence relation for $C_m$ (see Eq.~\ref{eq:ChebyRecur}) as well as the fact that $\bMEtilde(\bME)^{-1}\approx \bI$, $\bZ_k$ can be approximated as 
\begin{equation} \label{eq:residualapprox}
\begin{split}
    \bZ_k &\approx \bZtilde_k = \frac{2\sigma_{k}}{e} \bHE (\bMEtilde)^{-1}\bZtilde_{k-1} - \frac{2c\sigma_k}{e}\bZtilde_{k-1} \\
    & - \sigma_{k-1}\sigma_{k-2}\bZtilde_{k-2} +  \frac{2\sigma_{k}}{e} \bG \bLambda\,, 
\end{split}
\end{equation}
where $\bG = \bHE \bVin - \bME\bVin \bLambda$ is the eigen-residual of the approximate vectors and eigenvalues; $\bZtilde_0=\mathbf{0}$; and $\bZtilde_1=\frac{\sigma_1}{e}\bG$  . Subsequently, the Chebyshev subspace can be evaluated as $\bVout=(\bMEtilde)^{-1}\bZtilde_m + \bVin \bLambda$. Note that this residual-based reformulation of $\bVout$ does not require the evaluation of $(\bME)^{-1}$.  We refer to~\cite{kodali2025residual} for the details and the justification of why the residual Chebyshev filtering approach remains tolerant to inexact matrix-vector products.

We illustrate in Fig.~\ref{fig:chebycomparison} the efficacy of the residual Chebyshev filtering approach for copper (Cu) atom. As is evident, the residual Chebyshev filtering approach converges to tight tolerances close to machine precision. While this study demonstrated the robustness of the residual Chebyshev filtering approach for a single Cu atom, in Sec.~\ref {sec:results} we demonstrate its robustness on a wider range of material systems and system sizes.

\subsection{Electrostatics}
The solution of the Kohn-Sham problem also involves the solution of the electrostatic potential corresponding to the total charge density (cf.~Eq.~\ref{eq:poisson}). Here, we take advantage of the linearity of the problem and decompose the electrostatic potential into a part  coming from the superposition of atomic contributions and a remainder, i.e.
%
\begin{eqnarray}
\phi(\br) =\phi^{\text{at}}(\br) + \phirem(\br)\,.
\end{eqnarray}
In the above, $\phi^{\text{at}}(\br) = \sum_I \phi^{(I)}(\br) = \sum_I\frac{\rho^{(I)}(\br')+\bsmear^{(I)}(\br')}{|\br-\br'|}$ is the superposition of atomic $\phi^{(I)}$. Here $\rho^{(I)}$ is the density of an isolated atom of the same type as the $I$-th atom. Given the spherical symmetry of $\rho^{(I)}(\br)$, $\bsmear^{(I)}(\br)$, and $\phi^{(I)}(\br)$, they can be easily precomputed and stored for all atom types, using a 1D Poisson solve. The remainder potential, $\Delta \phi$, is obtained from the solution of a Poisson equation 
\begin{eqnarray}
    -\frac{1}{4 \pi} \nabla^{2} (\phirem(\br))=\Delta \rho(\br)
\end{eqnarray}
where $\Delta\rho(\br) = \rho(\br)-\sum_{I}\rho^{(I)}(\br)$.
%
We note that $\Delta\rho$ is a smoother field than $\rho$, and hence the corresponding potential is also smoother. Thus, the solution of the Poisson problem to compute $\phirem$ can be evaluated using a CFE basis without enrichment functions.  

\section{Results and Discussion}\label{sec:results}
We now present the accuracy and computational efficiency of the EFE basis. All DFT calculations are performed using norm-conserving ONCV PSPs from the \texttt{PseudoDojo} database \cite{van2018pseudodojo}, a temperature of 500~K for the Fermi-Dirac orbital occupancy, and the Anderson mixing~\cite{walker2011anderson} for the SCF iteration. All the simulations are performed on the CPU partition of the NERSC Perlmutter machine. Each Perlmutter CPU node consists of 2 AMD EPYC 7763 CPUs, with  64 cores per CPU and 512 GB of DDR4 memory per node. All the simulations used 64 MPI tasks per node, each task binding to two physical cores. The benchmark systems we consider in this study are non-periodic systems, with a subsequent work extending the implementation to periodic systems. 

\subsection{Accuracy of EFE basis}

We first present the systematic convergence of the EFE basis and provide a comparison with the CFE and planewave bases. For both EFE and CFE bases, systematic convergence can be achieved by decreasing the characteristic size of the finite-element mesh size ($h$), or increasing the finite-element polynomial order ($p$), or both. In the case of a planewave basis, systematic convergence can be obtained by increasing the planewave cutoff. In our study we use \texttt{DFT-FE}~\cite{motamarri2020dft, das2022dft} for CFE based calculations, and \texttt{Quantum Espresso (QE)}~\cite{Giannozzi2009quantum, giannozzi2017advanced} for planewave based calculations. In EFE and CFE calculations, we choose the simulation domain such that there is at least 24 a.u. vacuum, whereas in planewave calculations we use a vacuum of 12 a.u. The reference energies are obtained from a very refined CFE calculations using \texttt{DFT-FE}, with the $h\approx0.5$ a.u. and $p=6$, which provides discretization errors in energy of $\mathcal{O}(10^{-7})$ Ha/atom. 

Table~\ref{tab:aspirin} shows the results of the study for aspirin molecule ($\text{C}_9\text{H}_8\text{O}_4$). The discretization parameters for EFE and CFE basis are characterized in terms of the minimum and maximum finite-element sizes, denoted as $\hmin$ and $\hmax$, respectively, and the finite-element polynomial order ($p$).  The discretization for the planewave is characterized by the energy cutoff ($\Ecut$). Here, $\hmax$ is the finite-element size in the vacuum region, and $\hmin$ is the finite-element size near the atom as well as the key parameter that controls the discretization error in the energy. 
The error in the ground-state energy for various levels of discretization using EFE, CFE, and planewave basis is reported in Table~\ref{tab:aspirin}. As is evident, for all the basis sets, we obtain a systematic convergence. Notably, for the EFE basis, we obtain an accuracy of $10^{-4}$~Ha/atom---a typical desired discretization accuracy in DFT calculations---with $\sim3\times$ fewer DoFs/atom compared to the CFE basis  and with similar DoFs/atom as planewaves. We note that the gains of the EFE basis over the CFE basis are even greater ($5-7\times$ reduction in DoFs/atom) for larger systems, as will be discussed in Sec.~\ref {sec:efficiency}.

Table~\ref{tab:platinum} reports the same comparison as above but with a Pt icosahedral nanocluster with 13 atoms as the benchmark system. As with aspirin, we obtain the reference groundstate energy from a highly refined CFE basis calculation with the \texttt{DFT-FE} code. We, once again, observe a systematic convergence with all basis sets. We note that a discretization accuracy of $10^{-4}$~Ha/atom is attained in the EFE basis with $\sim3\times$ fewer basis functions than the CFE basis. However, the number of basis functions for the planewave basis to reach a similar accuracy for this system is $\sim3\times$ lower compared to the EFE basis.

\begin{table}[htbp]
\centering
\caption{Demonstration of systematic convergence of EFE basis and its comparison with CFE and planewave basis for the aspirin molecule. $\hmin, \hmax$, both in bohr, denote the smallest and largest finite-element size used in the CFE and EFE basis; $p$ is the finite-element polynomial order; $\Ecut$ is the planewave cutoff (in hartree); and DoFs/per atom is the number of basis functions per atom. Reference groundstate energy $E_0=-4.604113768$ Ha/atom  
}
\renewcommand{\arraystretch}{1.2}
\begin{tabular}{C{0.30\linewidth}|C{0.19\linewidth}|C{0.32\linewidth}}
\hline
 Parameters & DoFs/atom & Error (Ha/atom)   \\
\hline
\multicolumn{3}{l}{\textbf{EFE basis} ($\hmin, \hmax, p$)} \\
\hline
(1.25, 5, 4) & 7,672 & $2.87 \times 10^{-4}$ \\
(1.14, 4.55, 4) & 9,745 & $2.07 \times 10^{-4}$ \\
(0.89, 3.57, 4) & 19,456 & $6.36 \times 10^{-5}$  \\
(0.48, 3.85, 4) & 69,425 & $3.17 \times 10^{-6}$  \\
\hline
(2.5, 5, 5) & 7,069 & $1.83 \times 10^{-3}$ \\
(1.38, 5.56, 5) & 11,615 & $1.41 \times 10^{-4}$ \\
(1.14, 4.55, 5) & 18,686 & $5.60 \times 10^{-5}$ \\
(0.69, 5.56, 5) & 49,796 & $3.86 \times 10^{-6}$  \\
\hline
\multicolumn{3}{l}{\textbf{CFE basis} ($\hmin, \hmax, p$)} \\
\hline
(1.39, 5.56, 6) & 19,773 & $8.6 \times 10^{-4}$ \\
(1.14, 4.55, 6) & 31,876 & $8.7 \times 10^{-5}$ \\
(0.96, 3.85, 6) & 53,236 & $2.75 \times 10^{-5}$ \\
(0.69, 5.56, 6) & 85,151 & $1.1 \times 10^{-6}$  \\
\hline
\multicolumn{3}{l}{\textbf{Planewave basis}} \\
\hline
$\Ecut = 40$ Ha  & 9,438 & $1.96 \times 10^{-4}$  \\
$\Ecut = 50$ Ha  & 13,158 & $9.25 \times 10^{-5}$  \\
$\Ecut = 60$ Ha  & 17,315 & $7.29 \times 10^{-5}$  \\
$\Ecut = 150$ Ha  & 68,457 & $3.83 \times 10^{-6}$  \\
\hline
\end{tabular}
\label{tab:aspirin}
\end{table}

\begin{table}[htbp]
\centering
\caption{Demonstration of systematic convergence of EFE basis and its comparison with CFE and planewave basis for the  Pt-13 icosahedral nanocluster. $\hmin, \hmax$, both in bohr, denote the smallest and largest finite-element size used in the CFE and EFE basis; $p$ is the finite-element polynomial order; $\Ecut$ is the planewave cutoff (in hartree); and DoFs/atom is the number of basis functions per atom. Reference groundstate energy $E_0= -130.29617796$ Ha/atom.
}
\renewcommand{\arraystretch}{1.2}
\begin{tabular}{C{0.30\linewidth}|C{0.19\linewidth}|C{0.32\linewidth}}
\hline
 Parameters & DoFs/atom & $e$ (Ha/atom)   \\
\hline
\multicolumn{3}{l}{\textbf{EFE basis} ($\hmin, \hmax, p$)} \\
\hline
(1.25, 5, 4) & 17,821 & $4.10 \times 10^{-4}$  \\
(0.96, 3.85, 4) & 37,323 & $1.39 \times 10^{-4}$  \\
(0.48, 3.85, 4) & 194,449 & $1.69 \times 10^{-6}$  \\
\hline
(2.27, 4.55, 5) & 16,095 & $1.56 \times 10^{-3}$  \\
(1.25, 5, 5) & 34,138 & $1.20 \times 10^{-4}$  \\
(0.96, 3.85, 5) & 71,780 & $3.94 \times 10^{-5}$  \\
\hline
\multicolumn{3}{l}{\textbf{CFE basis} ($\hmin, \hmax, p$)} \\
\hline
(1.14, 4.55, 6) & 75,364 & $2.45 \times 10^{-4}$ \\
(1.04, 4.17, 6) & 97,217 & $9.90 \times 10^{-5}$ \\
(0.89, 3.57, 6) & 147,867 & $2.29 \times 10^{-5}$ \\
(0.69, 5.56, 6) & 228,619 & $1.957 \times 10^{-6}$ \\
\hline
\multicolumn{3}{l}{\textbf{Planewave basis}} \\
\hline
$\Ecut = 30$ Ha  & 8,141 & $8.12 \times 10^{-3}$  \\
$\Ecut = 40$ Ha  & 12,540 & $9.32 \times 10^{-5}$  \\
$\Ecut = 50$ Ha  & 17,533 & $4.55 \times 10^{-5}$  \\
$\Ecut = 150$ Ha  & 91,113 & $1.33 \times 10^{-6}$  \\
\hline
\end{tabular}
\label{tab:platinum}
\end{table}

\subsection{Computational efficiency of EFE basis} \label{sec:efficiency}
We now discuss the performance of the EFE basis using different materials systems with increasing system sizes reaching $\sim$39,000 electrons. We use copper (Cu) nanoparticles, sodium (Na) nanoparticles, and a DNA molecule as our benchmark materials systems to ascertain both the accuracy and efficiency of the EFE basis. We compare the performance of the EFE basis with the CFE basis, where the CFE simulations are, as before, performed using \texttt{DFT-FE}. We, particularly, assess the reduction in the DoFs attained by the EFE basis over the CFE basis. The discretization accuracy we aim for in this study is $10^{-4}$~Ha/atom. The reduction in DoFs also importantly affects the memory footprint. As a result, we also assess the minimum number of nodes required to execute a simulation using the EFE and CFE basis. Additionally, the coarser finite element basis in the EFE basis results in a smaller spectral width of the discrete Kohn-Sham Hamiltonian, which in turn reduces the Chebyshev polynomial order required in the residual Chebyshev filtering-based solution to the Kohn-Sham eigenvalue problem. As a result, we also compare the Chebyshev polynomial degree used in the EFE and CFE bases. Finally, we compare the overall computational time for the full ground-state calculation. 

The first materials system we consider is Cu nanoparticles of varying system sizes. We consider icosahedral clusters with different numbers of shells, namely Cu 2-shell (1,045 e-, where e- denotes electrons), Cu 4-shell (5,871 e-), Cu 5-shell (10,659 e-), and Cu 8-shell (39,083 e-). The nearest atom distance in these structures is 4.8 bohr. The reference ground-state energies for the Cu 2-shell, 4-shell, and 5-shell are obtained using \texttt{DFT-FE} using a refined FE basis with $\hmin=0.5$ a.u. and finite-element polynomial order ($p$) of 6. For the largest cluster (Cu 8-shell), given the high cost for the CFE basis, we only perform a few SCF iterations rather than a full ground-state calculation. As a result, the reported comparison for the Cu 8-shell is in terms of per SCF computational cost. In all the calculations, the CFE basis attains the desired accuracy of $10^{-4}$ Ha/atom with $\hmin=0.8$ a.u. and $p=6$. For the EFE basis, we attain the same accuracy with $\hmin=1.25$ a.u. and $p=5$. As noted in Table~\ref{tab:cucluster}, the EFE basis requires $\sim6\times$ fewer DoFs compared to the CFE basis. This reduction in DoFs also manifests as $5\times$ lower memory footprint (in terms of number of nodes) for the EFE basis. Further, as the spectral width for the discrete Hamiltonian is smaller for the EFE basis owing to the coarser discretization, EFE requires $2\times$ smaller Chebyshev polynomial degree compared to the CFE basis in the residual Chebyshev filtering. Overall, in terms of computational efficiency, the EFE basis provides a $\sim6\times$ speedup over the CFE basis, as implemented in a highly optimized code like \texttt{DFT-FE}.



For the second materials system, we consider Na nanoparticles constructed as icosahedral shells. In particular, we consider 4-shell and 5-shell nanoparticles with 2,781 e- and 5,049 e-, respectively. Table~\ref{tab:nacluster} reports the performance metrics, which show $~\sim6\times$ lower DoFs using the EFE basis in comparison to the CFE basis. The reduction in the memory footprint is $4\times$. Overall, the EFE basis is  $7-8\times$ faster compared to the CFE basis, in terms of the total computational time. 


As our last benchmark system, we consider a double‑stranded DNA fragment ($\mathrm{C_{310} H_{365} N_{113} O_{190} P_{30}}$) consisting of 1,008 atoms (3,460 e-). The geometry of the DNA molecule is taken from \cite{deng2020understanding}. As seen from the results in Table~\ref{tab:dna}, we attain a $5.4\times$ reduction in DoFs for the EFE basis. This results in a $4\times$ reduction in the memory footprint and a $\sim4.5\times$ improvement in computational efficiency.

These benchmark studies suggest that the EFE basis provides a substantial reduction in the number of basis functions in comparison to the CFE basis, which results in a substantial reduction in memory footprint and improvement in computational efficiency. The demonstrated benchmarks covering nanoparticles and an organic molecule suggest that the efficacy and efficiency of the EFE basis are independent of material systems and system size.

\begin{table*}[htbp]
\centering
\caption{Performance metrics of EFE and CFE basis for the copper (Cu) nanoparticles. CFE basis calculations are performed using \texttt{DFT-FE}. The Compute time is for the solution of the full groundstate solution, except for the Cu 8 shell, where we report the per SCF iteration time.  
}
\renewcommand{\arraystretch}{1.2}
\begin{tabular}{C{0.8in}|C{0.8in}|C{0.9in}|C{0.8in}|C{0.8in}|C{0.8in}}
\hline
Basis & DoFs/atom &  Error in Energy (Ha/atom) & Chebyshev Degree & Minimum nodes required & Compute time (Node-Hrs)\\
\hline
\multicolumn{6}{l}{\textbf{Cu 2-shell (55 atoms, 1,045 e-)}} \\
\hline 
EFE & 18,103 & $\quad 1.1\times 10^{-4}$ & 25 & 1 & 0.46\\
CFE  & 1,12,970 & $\quad 8\times 10^{-5}$ & 50 & 2 & 2.73 \\
\hline
\multicolumn{6}{l}{\textbf{Cu 4-shell (309 atoms, 5,871 e-)}} \\
\hline
EFE & 11,128 & $\quad 1.3\times 10^{-4}$ & 25 & 2 & 10.48\\
CFE & 69,696 & $\quad 9.8\times 10^{-5}$ & 50 & 10 & 59.31\\
\hline
\multicolumn{6}{l}{\textbf{Cu 5-shell (561 atoms, 10,659 e-)}} \\
\hline
EFE & 10,405 & $\quad 1.4\times 10^{-4}$ & 25 & 4 & 39.96 \\
CFE & 64,441 & $\quad 1.1\times 10^{-4}$ & 50 & 20 & 220.46\\
\hline
\multicolumn{6}{l}{\textbf{Cu 8-shell (2057 atoms, 39,083 e-)}} \\
\hline
EFE & 8,452 & \qquad -- & 25 & 28 & 12.29 (/SCF)\\
CFE & 55,539 & \qquad -- & 50 & 96 & 80.55 (/SCF)\\
\hline
\end{tabular}
\label{tab:cucluster}
\end{table*}

\begin{table*}[htbp]
\centering
\caption{Performance metrics of EFE and CFE basis for the sodium (Na) nanoparticles. CFE basis calculations are performed using \texttt{DFT-FE}.}
\renewcommand{\arraystretch}{1.2}
\begin{tabular}{C{0.8in}|C{0.8in}|C{0.9in}|C{0.8in}|C{0.8in}|C{0.8in}}
\hline
Basis & DoFs/atom & Error in Energy (Ha/atom) & Chebyshev Degree & Minimum nodes required & Compute time (Node-Hrs) \\
\hline
\multicolumn{6}{l}{\textbf{Na 4-shell (309 atoms, 2,781 e-)}} \\
\hline
EFE & 5,007 & $1.75\times 10^{-4}$ & 16 & 1 & 1.04\\
CFE & 31,327 & $1.8\times 10^{-4}$ & 50 & 4 & 9.18\\
\hline
\multicolumn{6}{l}{\textbf{Na 5-shell (561 atoms, 5,049 e-)}} \\
\hline
EFE & 4,624 & $1.84\times 10^{-4}$ & 16 & 2 & 5.71\\
CFE & 29,143 & $1.77\times 10^{-4}$ & 50 & 8 & 43.18\\
\hline
\end{tabular}
\label{tab:nacluster}
\end{table*}

\begin{table*}[htbp]
\centering
\caption{Performance metrics of EFE and CFE basis for the DNA molecule. CFE basis calculations are performed using \texttt{DFT-FE}.}
\label{tab:dna_cfe_EFE}
\renewcommand{\arraystretch}{1.2}
\renewcommand{\arraystretch}{1.2}
\begin{tabular}{C{0.8in}|C{0.8in}|C{0.9in}|C{0.8in}|C{0.8in}|C{0.8in}}
\hline
Basis & DoFs/atom & Error in Energy (Ha/atom) & Chebyshev Degree & Minimum nodes required &  Compute time (Node-Hrs)\\
\hline
\multicolumn{6}{l}{\textbf{DNA Molecule (1,008 atoms, 3,460 e-)}} \\
\hline
EFE & 3,125 & $7.6\times 10^{-5}$ & 16 & 2 & 8.77\\
CFE  & 16,906 & $1.48\times 10^{-4}$ & 30 & 8 & 39.19 \\
\hline
\end{tabular}
\label{tab:dna}
\end{table*}

\subsection{Scalability and timing breakdown using EFE basis}
\begin{figure}[htbp]
\includegraphics[width=8.7cm]{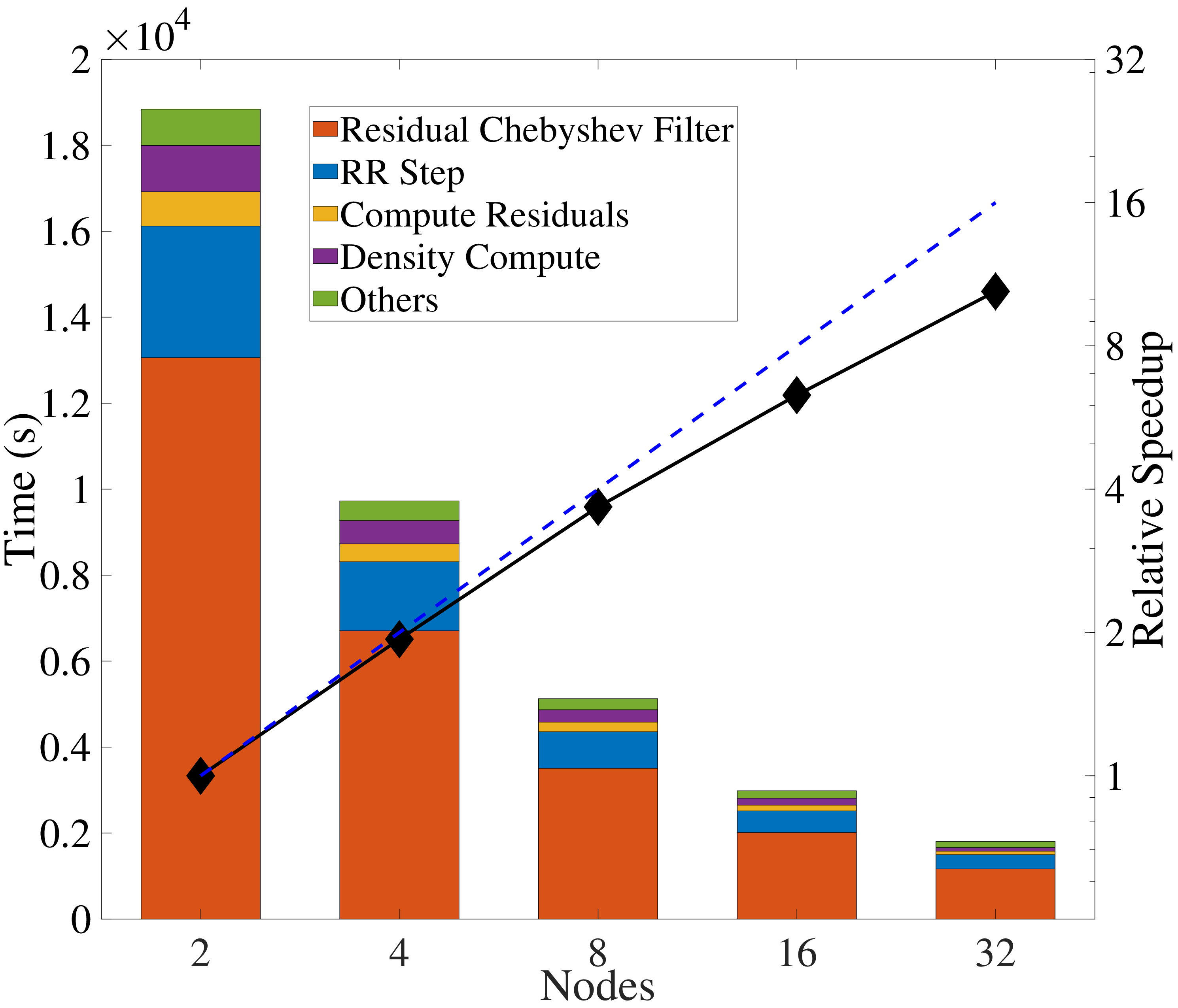}
\caption{\label{fig:epsart}  Breakdown of total wall-time into the various computational steps for Cu 4-shell nanoparticle (309 atoms, 5,871 e-)}
\label{cu4shellbreak}
\end{figure}

\begin{figure}[htbp]
\includegraphics[width=8.7cm]{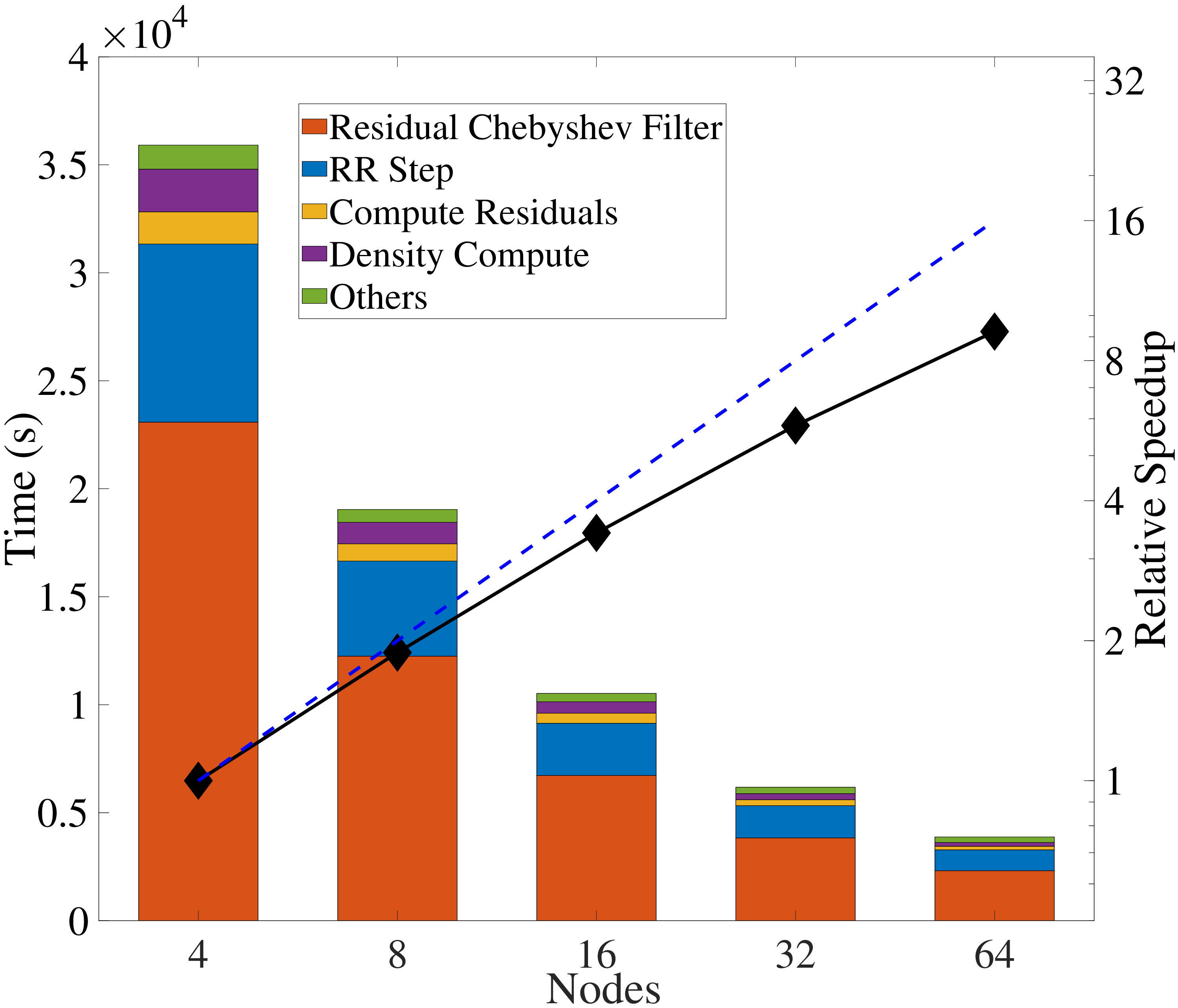}
\caption{\label{fig:epsart} Breakdown of total wall-time into the various computational steps for Cu 5-shell nanoparticle (561 atoms, 10,659 e-)}
\label{cu5shellbreak}
\end{figure}

We demonstrate the parallel scalability (strong scaling) using the EFE basis. Figures~\ref{cu4shellbreak} and~\ref{cu5shellbreak} show the parallel scalability of Cu-4 shell and Cu-5 shell systems. 

We obtain good scalability with a parallel efficiency of 65\% at $8 \times $ nodes for the Cu 4-shell system and 58\% efficiency for the Cu 5-shell system. We note that for the Cu 5-shell system on 64 nodes with 8192 MPI tasks, the average DoFs per MPI task is 1425, and hence, represents an extreme scaling regime. Compared to 4 nodes, the walltime reduces by $\sim10\times$ to 3893~secs at 64 nodes.



Figures~\ref{cu4shellbreak} and \ref{cu5shellbreak} also provide the breakdown of the strong scaling of various computational steps involved in computing the total ground state. The various steps, including construction of the EFE basis, construction of the discrete Hamiltonian, and Anderson-based density mixing, are provided under ``Other'' costs. The parallel scaling of these systems mainly is dependent on the scalability of the dominant computational steps: residual Chebyshev filtering and Rayleigh-Ritz, which make up around 70 \% and 16 \% for Cu 4-shell and 64 \% and 24 \% for Cu 5-shell, respectively. Chebyshev filtering is also the dominant cost for CFE basis~\cite{motamarri2020dft}. We  note that both residual Chebyshev filtering and Rayleigh-Ritz demonstrate an excellent parallel scalability, which results in an overall good strong scaling behavior. Although the computational cost of residual Chebyshev filtering scales quadratically, compared to the cubic scaling of the Rayleigh–Ritz procedure, its excellent parallel scalability enabled by the FE discretization remains critical for good parallel scalability for large system sizes.

Overall, we show that even with the addition of enrichment functions, the EFE basis still retains good parallel scalability, which is a result of the locality of the FE basis, the compactness of the enrichment functions, as well as an effective parallel implementation of the various algorithms. The parallel scalability provides the ability to conduct fast, large-scale DFT calculations using EFE basis. 



\section{Summary}
In this work, we presented an efficient approach for large-scale PSP DFT calculations using an orthogonalized enriched finite element (EFE) basis. The EFE basis is formed by augmenting the classical finite element (CFE) basis with compact atom-centered functions, called the enrichment functions. The key idea is to combine the efficiency of an atomic orbital basis with the completeness of the CFE basis. The enrichment functions are generated from the Kohn–Sham solutions to isolated atoms, which can be inexpensively pre-computed for the entire periodic table. To preserve locality and improve numerical conditioning, we truncated the enrichment functions with smooth cutoff functions. The inclusion of the enrichment functions enables an efficient means to represent the electronic fields in the vicinity of the nuclei, significantly reducing the need for a refined CFE basis. We also explicitly orthogonalized the enrichment functions with respect to the CFE basis to avoid any ill-conditioning in the resulting basis. The orthogonalization also leads to a block-diagonal form for the overlap matrix, simplifying the evaluation of its inverse.  

Further, we introduced several complementary measures to boost the overall computational efficiency and scalability of solving the generalized Kohn–Sham eigenvalue problem. First, we used spectral finite-elements along with reduced order Gauss–Lobatto–Legendre quadrature, the combination of which renders the classical–classical block of the overlap matrix diagonal. Second, we introduced an atom-block diagonal approximation to simplify the inverse of the enriched-enriched block of the overlap matrix. Third, to mitigate the effect of the above approximations to the inverse of the overlap matrix, we employed a residual Chebyshev filtering approach that is tolerant to an inexact inverse of the overlap matrix. 

We demonstrated the accuracy and efficiency of the EFE basis using different benchmark systems, including aspirin molecule, platinum, copper, and sodium nanoclusters, and DNA molecule, with the largest system containing 39,083 electrons. We showed that the EFE basis requires $5-7\times$ fewer DoFs/atom compared to the CFE basis, and hence, further bridges the gap between planewave and FE basis. The EFE attains a $5-9\times$ speedup over the CFE, along with a $4-5\times$ reduction in the requisite computational memory. Lastly, the EFE basis, despite the inclusion of enrichment functions, shows good parallel scalability even at an $16$-fold increase in the number of CPUs.

Overall, the proposed EFE framework provides a computationally efficient, systematically convergent, and massively parallelizable method for large-scale PSP DFT calculations. Although our results are limited to non-periodic systems, we intend to extend it to periodic and semi-periodic systems as well. The EFE basis, along with configurational forces~\cite{Motamarri2018, Rufus2022}, can offer a fast method for \textit{ab initio} molecular dynamics and is part of our immediate future effort. We expect the EFE basis to be amenable to substantial acceleration on hybrid CPU-GPU architectures and is being actively pursued. Finally, the EFE also offers good promise for accurate and efficient PSP time-dependent density functional theory (TDDFT) calculations~\cite{Kanungo2019TDDFT, kanungo2023TDDFT}.  


\acknowledgements
We gratefully acknowledge the support from the Department of Energy, Office of Basic Energy Sciences, Award No. DE-SC0008637, that funded this work.  This research used resources of the National Energy Research Scientific Computing Center, a DOE Office of Science User Facility supported by the Office of Science of the U.S. Department of Energy under Contract No. DE-AC02-05CH11231. 

\bibliography{apssamp}

@article{pask2017partition,
  title={Partition of unity finite element method for quantum mechanical materials calculations},
  author={Pask, John E and Sukumar, Natarajan},
  journal={Extreme Mechanics Letters},
  volume={11},
  pages={8--17},
  year={2017},
  publisher={Elsevier}
}

@article{hohenberg1964inhomogeneous,
  title={Inhomogeneous electron gas},
  author={Hohenberg, Pierre and Kohn, Walter},
  journal={Physical review},
  volume={136},
  number={3B},
  pages={B864},
  year={1964},
  publisher={APS}
}

@article{kohn1965self,
  title={Self-consistent equations including exchange and correlation effects},
  author={Kohn, Walter and Sham, Lu Jeu},
  journal={Physical review},
  volume={140},
  number={4A},
  pages={A1133},
  year={1965},
  publisher={APS}
}

@book{martin2020electronic,
  title={Electronic structure: basic theory and practical methods},
  author={Martin, Richard M},
  year={2020},
  publisher={Cambridge university press}
}

@article{cole2016applications,
  title={Applications of large-scale density functional theory in biology},
  author={Cole, Daniel J and Hine, Nicholas DM},
  journal={Journal of Physics: Condensed Matter},
  volume={28},
  number={39},
  pages={393001},
  year={2016},
  publisher={IOP Publishing}
}

@article{das2017electronic,
  title={Electronic structure study of screw dislocation core energetics in Aluminum and core energetics informed forces in a dislocation aggregate},
  author={Das, Sambit and Gavini, Vikram},
  journal={Journal of the Mechanics and Physics of Solids},
  volume={104},
  pages={115--143},
  year={2017},
  publisher={Elsevier}
}

@article{ismail2000ab,
  title={Ab initio study of screw dislocations in {Mo} and {Ta}: a new picture of plasticity in bcc transition metals},
  author={Ismail-Beigi, Sohrab and Arias, TA},
  journal={Physical Review Letters},
  volume={84},
  number={7},
  pages={1499},
  year={2000},
  publisher={APS}
}

@article{shin2013possible,
  title={Possible origin of the discrepancy in {Peierls} stresses of fcc metals: First-principles simulations of dislocation mobility in aluminum},
  author={Shin, Ilgyou and Carter, Emily A},
  journal={Physical Review B—Condensed Matter and Materials Physics},
  volume={88},
  number={6},
  pages={064106},
  year={2013},
  publisher={APS}
}

@article{rodney2017ab,
  title={Ab initio modeling of dislocation core properties in metals and semiconductors},
  author={Rodney, David and Ventelon, L and Clouet, E and Pizzagalli, Laurent and Willaime, F},
  journal={Acta Materialia},
  volume={124},
  pages={633--659},
  year={2017},
  publisher={Elsevier}
}

@article{dive2018molecular,
  title={Molecular dynamics modeling of the structure and {Na+}-ion transport in {Na2S+} {SiS2} glassy electrolytes},
  author={Dive, A and Benmore, C and Wilding, M and Martin, SW and Beckman, S and Banerjee, Soumik},
  journal={The Journal of Physical Chemistry B},
  volume={122},
  number={30},
  pages={7597--7608},
  year={2018},
  publisher={ACS Publications}
}

@article{schwerdtfeger2011pseudopotential,
  title={The pseudopotential approximation in electronic structure theory},
  author={Schwerdtfeger, Peter},
  journal={ChemPhysChem},
  volume={12},
  number={17},
  pages={3143--3155},
  year={2011},
  publisher={Wiley Online Library}
}

@article{vanderbilt1990soft,
  title={Soft self-consistent pseudopotentials in a generalized eigenvalue formalism},
  author={Vanderbilt, David},
  journal={Physical review B},
  volume={41},
  number={11},
  pages={7892},
  year={1990},
  publisher={APS}
}

@article{blochl1994projector,
  title={Projector augmented-wave method},
  author={Bl{\"o}chl, Peter E},
  journal={Physical review B},
  volume={50},
  number={24},
  pages={17953},
  year={1994},
  publisher={APS}
}

@article{chelikowsky1994finite,
  title={Finite-difference-pseudopotential method: Electronic structure calculations without a basis},
  author={Chelikowsky, James R and Troullier, N and Saad, Yousef},
  journal={Physical review letters},
  volume={72},
  number={8},
  pages={1240},
  year={1994},
  publisher={APS}
}

@article{das2022dft,
  title={{DFT-FE} 1.0: A massively parallel hybrid {CPU-GPU} density functional theory code using finite-element discretization},
  author={Das, Sambit and Motamarri, Phani and Subramanian, Vishal and Rogers, David M and Gavini, Vikram},
  journal={Computer Physics Communications},
  volume={280},
  pages={108473},
  year={2022},
  publisher={Elsevier}
}

@article{kanungo2017large,
  title={Large-scale all-electron density functional theory calculations using an enriched finite-element basis},
  author={Kanungo, Bikash and Gavini, Vikram},
  journal={Physical Review B},
  volume={95},
  number={3},
  pages={035112},
  year={2017},
  publisher={APS}
}

@article{rufus2021fast,
  title={Fast and robust all-electron density functional theory calculations in solids using orthogonalized enriched finite elements},
  author={Rufus, Nelson D and Kanungo, Bikash and Gavini, Vikram},
  journal={Physical Review B},
  volume={104},
  number={8},
  pages={085112},
  year={2021},
  publisher={APS}
}

@article{motamarri2020dft,
  title={{DFT-FE:} {A} massively parallel adaptive finite-element code for large-scale density functional theory calculations},
  author={Motamarri, Phani and Das, Sambit and Rudraraju, Shiva and Ghosh, Krishnendu and Davydov, Denis and Gavini, Vikram},
  journal={Computer Physics Communications},
  volume={246},
  pages={106853},
  year={2020},
  publisher={Elsevier}
}

@article{motamarri2013higher,
  title={Higher-order adaptive finite-element methods for {Kohn--Sham} density functional theory},
  author={Motamarri, Phani and Nowak, Michael R and Leiter, Kenneth and Knap, Jaroslaw and Gavini, Vikram},
  journal={Journal of Computational Physics},
  volume={253},
  pages={308--343},
  year={2013},
  publisher={Elsevier}
}

@inproceedings{das2023large,
  title={Large-scale materials modeling at quantum accuracy: Ab initio simulations of quasicrystals and interacting extended defects in metallic alloys},
  author={Das, Sambit and Kanungo, Bikash and Subramanian, Vishal and Panigrahi, Gourab and Motamarri, Phani and Rogers, David and Zimmerman, Paul and Gavini, Vikram},
  booktitle={Proceedings of the International Conference for High Performance Computing, Networking, Storage and Analysis},
  pages={1--12},
  year={2023}
}

@article{subramanian2024tucker,
  title={Tucker Tensor Approach for Accelerating Fock Exchange Computations in a Real-Space Finite-Element Discretization of Generalized {Kohn--Sham} Density Functional Theory},
  author={Subramanian, Vishal and Das, Sambit and Gavini, Vikram},
  journal={Journal of Chemical Theory and Computation},
  volume={20},
  number={9},
  pages={3566--3579},
  year={2024},
  publisher={ACS Publications}
}

@article{kanungo2019exact,
  title={Exact exchange-correlation potentials from ground-state electron densities},
  author={Kanungo, Bikash and Zimmerman, Paul M and Gavini, Vikram},
  journal={Nature communications},
  volume={10},
  number={1},
  pages={4497},
  year={2019},
  publisher={Nature Publishing Group UK London}
}

@article{baek2025quasicrystal,
  title={Quasicrystal stability and nucleation kinetics from density functional theory},
  author={Baek, Woohyeon and Das, Sambit and Tan, Shibo and Gavini, Vikram and Sun, Wenhao},
  journal={Nature Physics},
  pages={1--8},
  year={2025},
  publisher={Nature Publishing Group UK London}
}

@article{zhuravel2020backbone,
  title={Backbone charge transport in double-stranded {DNA}},
  author={Zhuravel, Roman and Huang, Haichao and Polycarpou, Georgia and Polydorides, Savvas and Motamarri, Phani and Katrivas, Liat and Rotem, Dvir and Sperling, Joseph and Zotti, Linda A and Kotlyar, Alexander B and others},
  journal={Nature nanotechnology},
  volume={15},
  number={10},
  pages={836--840},
  year={2020},
  publisher={Nature Publishing Group UK London}
}

@article{ghosh2019all,
  title={All-electron density functional calculations for electron and nuclear spin interactions in molecules and solids},
  author={Ghosh, Krishnendu and Ma, He and Gavini, Vikram and Galli, Giulia},
  journal={Physical Review Materials},
  volume={3},
  number={4},
  pages={043801},
  year={2019},
  publisher={APS}
}

@article{pask2012linear,
  title={Linear scaling solution of the all-electron coulomb problem insolids},
  author={Pask, JE and Sukumar, N and Mousavi, SE},
  journal={International Journal for Multiscale Computational Engineering},
  volume={10},
  number={1},
  year={2012},
  publisher={Begel House Inc.}
}

@article{giannozzi2017advanced,
  title={Advanced capabilities for materials modelling with {Quantum ESPRESSO}},
  author={Giannozzi, Paolo and Andreussi, Oliviero and Brumme, Thomas and Bunau, Oana and Nardelli, M Buongiorno and Calandra, Matteo and Car, Roberto and Cavazzoni, Carlo and Ceresoli, Davide and Cococcioni, Matteo and others},
  journal={Journal of physics: Condensed matter},
  volume={29},
  number={46},
  pages={465901},
  year={2017},
  publisher={IOP Publishing}
}

@article{van2018pseudodojo,
  title={The {PseudoDojo}: Training and grading a 85 element optimized norm-conserving pseudopotential table},
  author={Van Setten, Michiel J and Giantomassi, Matteo and Bousquet, Eric and Verstraete, Matthieu J and Hamann, Don R and Gonze, Xavier and Rignanese, G-M},
  journal={Computer Physics Communications},
  volume={226},
  pages={39--54},
  year={2018},
  publisher={Elsevier}
}

@article{kodali2025residual,
  title={Residual-based Chebyshev filtered subspace iteration for sparse Hermitian eigenvalue problems tolerant to inexact matrix-vector products},
  author={Kodali, Nikhil and Ramakrishnan, Kartick and Motamarri, Phani},
  journal={Computer Physics Communications},
  volume={327},
  pages={110239},
  year={2026}
}

@article{zhou2006parallel,
  title={Parallel self-consistent-field calculations via Chebyshev-filtered subspace acceleration},
  author={Zhou, Yunkai and Saad, Yousef and Tiago, Murilo L and Chelikowsky, James R},
  journal={Physical Review E—Statistical, Nonlinear, and Soft Matter Physics},
  volume={74},
  number={6},
  pages={066704},
  year={2006},
  publisher={APS}
}

@article{zhou2014chebyshev,
  title={Chebyshev-filtered subspace iteration method free of sparse diagonalization for solving the {Kohn--Sham} equation},
  author={Zhou, Yunkai and Chelikowsky, James R and Saad, Yousef},
  journal={Journal of Computational Physics},
  volume={274},
  pages={770--782},
  year={2014},
  publisher={Elsevier}
}

@article{walker2011anderson,
  title={Anderson acceleration for fixed-point iterations},
  author={Walker, Homer F and Ni, Peng},
  journal={SIAM Journal on Numerical Analysis},
  volume={49},
  number={4},
  pages={1715--1735},
  year={2011},
  publisher={SIAM}
}

@article{deng2020understanding,
  title={Understanding atomic bonding and electronic distributions of a {DNA} molecule using {DFT} calculation and {BOLS-BC} model},
  author={Deng, Anlin and Li, Hanze and Bo, Maolin and Huang, ZhongKai and Li, Lei and Yao, Chuang and Li, Fengqin},
  journal={Biochemistry and Biophysics Reports},
  volume={24},
  pages={100804},
  year={2020},
  publisher={Elsevier}
}

@article{Pask1999,
  title = {Real-space local polynomial basis for solid-state electronic-structure calculations: A finite-element approach},
  author = {Pask, J. E. and Klein, B. M. and Fong, C. Y. and Sterne, P. A.},
  journal = {Phys. Rev. B},
  volume = {59},
  issue = {19},
  pages = {12352--12358},
  numpages = {0},
  year = {1999},
  month = {May},
  publisher = {American Physical Society},
  doi = {10.1103/PhysRevB.59.12352},
  url = {http://link.aps.org/doi/10.1103/PhysRevB.59.12352}
}

@article{pask2001finite,
  title={Finite-element methods in electronic-structure theory},
  author={Pask, John E and Klein, Barry M and Sterne, Philip A and Fong, Chingyao Y},
  journal={Computer Physics Communications},
  volume={135},
  number={1},
  pages={1--34},
  year={2001},
  publisher={Elsevier}
}

@article{hamann2013optimized,
  title={Optimized norm-conserving Vanderbilt pseudopotentials},
  author={Hamann, Don R},
  journal={Physical Review B—Condensed Matter and Materials Physics},
  volume={88},
  number={8},
  pages={085117},
  year={2013},
  publisher={APS}
}

@article{perdew1992accurate,
  title={Accurate and simple analytic representation of the electron-gas correlation energy},
  author={Perdew, John P and Wang, Yue},
  journal={Physical review B},
  volume={45},
  number={23},
  pages={13244},
  year={1992},
  publisher={APS}
}

@ARTICLE{Pask2005,
author={Pask, J. E. and Sterne, P. A.},
title={Finite element methods in ab initio electronic structure calculations},
journal={Modelling and Simulation in Materials Science and Engineering},
year={2005},
volume={13},
number={3},
pages={R71-R96},
doi={10.1088/0965-0393/13/3/R01},
url={https://www.scopus.com/inward/record.uri?eid=2-s2.0-24144475862&partnerID=40&md5=756ec01428ae4ab7cc7091b760e2b9d8},
document_type={Review},
source={Scopus},
}

@article{Motamarri2012,
title = "Higher-order adaptive finite-element methods for orbital-free density functional theory ",
journal = "Journal of Computational Physics ",
volume = "231",
number = "20",
pages = "6596 - 6621",
year = "2012",
note = "",
issn = "0021-9991",
doi = "http://dx.doi.org/10.1016/j.jcp.2012.04.036",
url = "http://www.sciencedirect.com/science/article/pii/S0021999112002185",
author = "Phani Motamarri and Mrinal Iyer and Jaroslaw Knap and Vikram Gavini"
}

@article{Suryanarayana2010,
title = "Non-periodic finite-element formulation of {Kohn–Sham} density functional theory ",
journal = "Journal of the Mechanics and Physics of Solids ",
volume = "58",
number = "2",
pages = "256 - 280",
year = "2010",
note = "",
issn = "0022-5096",
doi = "http://dx.doi.org/10.1016/j.jmps.2009.10.002",
url = "http://www.sciencedirect.com/science/article/pii/S0022509609001409",
author = "Phanish Suryanarayana and Vikram Gavini and Thomas Blesgen and Kaushik Bhattacharya and Michael Ortiz"
}

@article{zhou2006self,
  title={Self-consistent-field calculations using Chebyshev-filtered subspace iteration},
  author={Zhou, Yunkai and Saad, Yousef and Tiago, Murilo L and Chelikowsky, James R},
  journal={Journal of Computational Physics},
  volume={219},
  number={1},
  pages={172--184},
  year={2006},
  publisher={Elsevier}
}

@article{Cohen2012Challenges,
  title={Challenges for density functional theory},
  author={Cohen, Aron J and Mori-S{\'a}nchez, Paula and Yang, Weitao},
  journal={Chemical reviews},
  volume={112},
  number={1},
  pages={289--320},
  year={2012},
  publisher={ACS Publications}
}

@article{Burke2012Perspective,
  title={Perspective on density functional theory},
  author={Burke, Kieron},
  journal={The Journal of chemical physics},
  volume={136},
  number={15},
  year={2012},
  publisher={AIP Publishing}
}

@article{Mardirossian2017Thirty,
  title={Thirty years of density functional theory in computational chemistry: an overview and extensive assessment of 200 density functionals},
  author={Mardirossian, Narbe and Head-Gordon, Martin},
  journal={Molecular physics},
  volume={115},
  number={19},
  pages={2315--2372},
  year={2017},
  publisher={Taylor \& Francis}
}

@article{Jones2015Density,
  title={Density functional theory: Its origins, rise to prominence, and future},
  author={Jones, Robert O},
  journal={Reviews of modern physics},
  volume={87},
  number={3},
  pages={897--923},
  year={2015},
  publisher={APS}
}

@article{Teale2022DFT,
  title={{DFT} exchange: sharing perspectives on the workhorse of quantum chemistry and materials science},
  author={Teale, Andrew M and Helgaker, Trygve and Savin, Andreas and Adamo, Carlo and Aradi, B{\'a}lint and Arbuznikov, Alexei V and Ayers, Paul W and Baerends, Evert Jan and Barone, Vincenzo and Calaminici, Patrizia and others},
  journal={Physical chemistry chemical physics},
  volume={24},
  number={47},
  pages={28700--28781},
  year={2022},
  publisher={Royal Society of Chemistry}
}

@article{Chelikowsky2000PSP,
  title={The pseudopotential-density functional method applied to nanostructures},
  author={Chelikowsky, James R},
  journal={Journal of Physics D: Applied Physics},
  volume={33},
  number={8},
  pages={R33--R50},
  year={2000}
}

@article{Pickett1989PSP,
  title={Pseudopotential methods in condensed matter applications},
  author={Pickett, Warren E},
  journal={Computer Physics Reports},
  volume={9},
  number={3},
  pages={115--197},
  year={1989},
  publisher={Elsevier}
}

@article{Carr2020electronic,
  title={Electronic-structure methods for twisted moir{\'e} layers},
  author={Carr, Stephen and Fang, Shiang and Kaxiras, Efthimios},
  journal={Nature Reviews Materials},
  volume={5},
  number={10},
  pages={748--763},
  year={2020},
  publisher={Nature Publishing Group UK London}
}

@article{Soler2002siesta,
  title={The {SIESTA} method for ab initio order-{N} materials simulation},
  author={Soler, Jos{\'e} M and Artacho, Emilio and Gale, Julian D and Garc{\'\i}a, Alberto and Junquera, Javier and Ordej{\'o}n, Pablo and S{\'a}nchez-Portal, Daniel},
  journal={Journal of physics: Condensed matter},
  volume={14},
  number={11},
  pages={2745--2779},
  year={2002}
}

@article{Kuhne2020cp2k,
  title={{CP2K}: An electronic structure and molecular dynamics software package-Quickstep: Efficient and accurate electronic structure calculations},
  author={K{\"u}hne, Thomas D and Iannuzzi, Marcella and Del Ben, Mauro and Rybkin, Vladimir V and Seewald, Patrick and Stein, Frederick and Laino, Teodoro and Khaliullin, Rustam Z and Sch{\"u}tt, Ole and Schiffmann, Florian and others},
  journal={The Journal of Chemical Physics},
  volume={152},
  number={19},
  year={2020},
  publisher={AIP Publishing}
}

@article{Jensen2013atomic,
  title={Atomic orbital basis sets},
  author={Jensen, Frank},
  journal={Wiley Interdisciplinary Reviews: Computational Molecular Science},
  volume={3},
  number={3},
  pages={273--295},
  year={2013},
  publisher={Wiley Online Library}
}

@article{Payne1992,
  title = {Iterative minimization techniques for ab initio total-energy calculations: molecular dynamics and conjugate gradients},
  author = {Payne, M. C. and Teter, M. P. and Allan, D. C. and Arias, T. A. and Joannopoulos, J. D.},
  journal = {Rev. Mod. Phys.},
  volume = {64},
  issue = {4},
  pages = {1045--1097},
  numpages = {0},
  year = {1992},
  month = {Oct},
  publisher = {American Physical Society},
  doi = {10.1103/RevModPhys.64.1045},
  url = {https://link.aps.org/doi/10.1103/RevModPhys.64.1045}
}

@article{Kresse1996efficient,
  title={Efficient iterative schemes for ab initio total-energy calculations using a plane-wave basis set},
  author={Kresse, Georg and Furthm{\"u}ller, J{\"u}rgen},
  journal={Physical review B},
  volume={54},
  number={16},
  pages={11169},
  year={1996},
  publisher={APS}
}

@article{Giannozzi2009quantum,
  title={{QUANTUM ESPRESSO}: a modular and open-source software project for quantum simulations of materials},
  author={Giannozzi, Paolo and Baroni, Stefano and Bonini, Nicola and Calandra, Matteo and Car, Roberto and Cavazzoni, Carlo and Ceresoli, Davide and Chiarotti, Guido L and Cococcioni, Matteo and Dabo, Ismaila and others},
  journal={Journal of physics: Condensed matter},
  volume={21},
  number={39},
  pages={395502},
  year={2009}
}

@article{Gonze2002first,
  title={First-principles computation of material properties: the {ABINIT} software project},
  author={Gonze, Xavier and Beuken, J-M and Caracas, Razvan and Detraux, F and Fuchs, M and Rignanese, G-M and Sindic, Luc and Verstraete, Matthieu and Zerah, G and Jollet, F and others},
  journal={Computational Materials Science},
  volume={25},
  number={3},
  pages={478--492},
  year={2002},
  publisher={Elsevier}
}

@article{Bowler2006recent,
  title={Recent progress with large-scale ab initio calculations: the CONQUEST code},
  author={Bowler, DR and Choudhury, R and Gillan, MJ and Miyazaki, T},
  journal={Physica Status Solidi (b)},
  volume={243},
  number={5},
  pages={989--1000},
  year={2006},
  publisher={Wiley Online Library}
}

@article{Kronik2006parsec,
  title={{PARSEC}--the pseudopotential algorithm for real-space electronic structure calculations: recent advances and novel applications to nano-structures},
  author={Kronik, Leeor and Makmal, Adi and Tiago, Murilo L and Alemany, MMG and Jain, Manish and Huang, Xiangyang and Saad, Yousef and Chelikowsky, James R},
  journal={Physica Status Solidi (b)},
  volume={243},
  number={5},
  pages={1063--1079},
  year={2006},
  publisher={Wiley Online Library}
}

@article{Ghosh2017sparc,
  title={{SPARC}: Accurate and efficient finite-difference formulation and parallel implementation of density functional theory: Isolated clusters},
  author={Ghosh, Swarnava and Suryanarayana, Phanish},
  journal={Computer Physics Communications},
  volume={212},
  pages={189--204},
  year={2017},
  publisher={Elsevier}
}

@article{Tsuchida1995electronic,
  title={Electronic-structure calculations based on the finite-element method},
  author={Tsuchida, Eiji and Tsukada, Masaru},
  journal={Physical Review B},
  volume={52},
  number={8},
  pages={5573},
  year={1995},
  publisher={APS}
}

@article{White1989finite,
  title={Finite-element method for electronic structure},
  author={White, Steven R and Wilkins, John W and Teter, Michael P},
  journal={Physical Review B},
  volume={39},
  number={9},
  pages={5819},
  year={1989},
  publisher={APS}
}

@inproceedings{Das2019fast,
  title={Fast, scalable and accurate finite-element based ab initio calculations using mixed precision computing: 46 {PFLOPS} simulation of a metallic dislocation system},
  author={Das, Sambit and Motamarri, Phani and Gavini, Vikram and Turcksin, Bruno and Li, Ying Wai and Leback, Brent},
  booktitle={Proceedings of the international conference for high performance computing, networking, storage and analysis},
  pages={1--11},
  year={2019}
}

@article{hamann1979norm,
  title={Norm-conserving pseudopotentials},
  author={Hamann, DR and Schl{\"u}ter, M and Chiang, C},
  journal={Phys. Rev. Lett.},
  volume={43},
  number={20},
  pages={1494},
  year={1979},
  publisher={APS}
}

@article{goedecker1996separable,
  title={Separable dual-space Gaussian pseudopotentials},
  author={Goedecker, Stefan and Teter, Michael and Hutter, J{\"u}rg},
  journal={Physical Review B},
  volume={54},
  number={3},
  pages={1703},
  year={1996},
  publisher={APS}
}

@article{Motamarri2018,
  title = {Configurational forces in electronic structure calculations using {Kohn-Sham} density functional theory},
  author = {Motamarri, Phani and Gavini, Vikram},
  journal = {Phys. Rev. B},
  volume = {97},
  issue = {16},
  pages = {165132},
  numpages = {30},
  year = {2018},
  month = {Apr},
  publisher = {American Physical Society},
  doi = {10.1103/PhysRevB.97.165132},
  url = {https://link.aps.org/doi/10.1103/PhysRevB.97.165132}
}

@article{Rufus2022,
  title = {Ionic forces and stress tensor in all-electron density functional theory calculations using an enriched finite-element basis},
  author = {Rufus, Nelson D. and Gavini, Vikram},
  journal = {Phys. Rev. B},
  volume = {106},
  issue = {8},
  pages = {085108},
  numpages = {17},
  year = {2022},
  month = {Aug},
  publisher = {American Physical Society},
  doi = {10.1103/PhysRevB.106.085108},
  url = {https://link.aps.org/doi/10.1103/PhysRevB.106.085108}
}

@article{Kanungo2019TDDFT,
  title = {Real time time-dependent density functional theory using higher order finite-element methods},
  author = {Kanungo, Bikash and Gavini, Vikram},
  journal = {Phys. Rev. B},
  volume = {100},
  issue = {11},
  pages = {115148},
  numpages = {26},
  year = {2019},
  month = {Sep},
  publisher = {American Physical Society},
  doi = {10.1103/PhysRevB.100.115148},
  url = {https://link.aps.org/doi/10.1103/PhysRevB.100.115148}
}

@article{kanungo2023TDDFT,
  title={Efficient all-electron time-dependent density functional theory calculations using an enriched finite element basis},
  author={Kanungo, Bikash and Rufus, Nelson D and Gavini, Vikram},
  journal={Journal of Chemical Theory and Computation},
  volume={19},
  number={3},
  pages={978--991},
  year={2023},
  publisher={ACS Publications}
}

@article{Melenk1996,
title = "The partition of unity finite element method: Basic theory and applications ",
journal = "Computer Methods in Applied Mechanics and Engineering ",
volume = "139",
number = "1–4",
pages = "289 - 314",
year = "1996",
note = "",
issn = "0045-7825",
doi = "http://dx.doi.org/10.1016/S0045-7825(96)01087-0",
url = "http://www.sciencedirect.com/science/article/pii/S0045782596010870",
author = "J. M. Melenk and I. Babu{\v{s}}ka"
}

@article {Babuska1997,
author = {Babu{\v{s}}ka, I. and Melenk, J. M.},
title = {THE PARTITION OF UNITY METHOD},
journal = {International Journal for Numerical Methods in Engineering},
volume = {40},
number = {4},
publisher = {John Wiley & Sons, Ltd},
issn = {1097-0207},
url = {http://dx.doi.org/10.1002/(SICI)1097-0207(19970228)40:4<727::AID-NME86>3.0.CO;2-N},
doi = {10.1002/(SICI)1097-0207(19970228)40:4<727::AID-NME86>3.0.CO;2-N},
pages = {727--758},
year = {1997},
}

@article{das2026,
title = {Intrinsic ductility enhancement in {Mg} alloys elucidated via large-scale ab-initio calculations},
journal = {Acta Materialia},
volume = {315},
pages = {122377},
year = {2026},
issn = {1359-6454},
doi = {https://doi.org/10.1016/j.actamat.2026.122377},
url = {https://www.sciencedirect.com/science/article/pii/S1359645426004787},
author = {Sambit Das and Vikram Gavini}
}

@article{ghosh2021,
title = {Spin–spin interactions in defects in solids from mixed all-electron and pseudopotential first-principles calculations},
journal = {npj Comput Mater},
volume = {7},
pages = {123},
year = {2021},
author = {Krishnendu Ghosh and He Ma and Mykyta Onizhuk and Vikram Gavini and Giulia Galli}
}

\end{document}